## Research Article

# Probing the soft X-ray properties and multi-wavelength variability of SN2023ixf and its progenitor

Sonja Panjkov[1] , Katie Auchettl[1,2] , Benjamin J. Shappee[3] , Aaron Do[3] , Laura Lopez[4,5,6] , and John F. Beacom[4,5,7]

[1]OzGrav, School of Physics, The University of Melbourne, Parkville, VIC, Australia, [2]Department of Astronomy and Astrophysics, University of California, Santa Cruz, CA, USA, [3]Institute for Astronomy, University of Hawai'i, Honolulu, HI, USA, [4]Department of Astronomy, Ohio State University, Columbus, OH, USA, [5]Center for Cosmology and AstroParticle Physics (CCAPP), Ohio State University, Columbus, OH, USA, [6]Flatiron Institute, Center for Computational Astrophysics, New York, NY, USA and [7]Department of Physics, Ohio State University, Columbus, OH, USA

**Abstract**

We present a detailed analysis of nearly two decades of optical/UV and X-ray data to study the multi-wavelength pre-explosion properties and post-explosion X-ray properties of nearby SN2023ixf located in M101. We find no evidence of precursor activity in the optical to UV down to a luminosity of $\lesssim 1.0 \times 10^5 \, L_\odot$, while X-ray observations covering nearly 18 yr prior to explosion show no evidence of luminous precursor X-ray emission down to an absorbed 0.3–10.0 keV X-ray luminosity of $\sim 6 \times 10^{36}$ erg s$^{-1}$. Extensive *Swift* observations taken post-explosion did not detect soft X-ray emission from SN2023ixf within the first $\sim 3.3$ days after first light, which suggests a mass-loss rate for the progenitor of $\lesssim 5 \times 10^{-4} \, M_\odot$ yr$^{-1}$ or a radius of $\lesssim 4 \times 10^{15}$ cm for the circumstellar material. Our analysis also suggests that if the progenitor underwent a mass-loss episode, this had to occur $> 0.5$–1.5 yr prior to explosion, consistent with previous estimates. *Swift* detected soft X-rays from SN2023ixf $\sim 4.25$ days after first light, and it rose to a peak luminosity of $\sim 10^{39}$ erg s$^{-1}$ after 10 days and has maintained this luminosity for nearly 50 days post first light. This peak luminosity is lower than expected, given the interaction that SN2023ixf is interacting with dense material. However, this might be a natural consequence of an asymmetric circumstellar medium. X-ray spectra derived from merging all *Swift* observations over the first 50 days are best described by a two-component bremsstrahlung model consisting of a heavily absorbed and hotter component similar to that found using *NuSTAR*, and a less-absorbed, cooler component. We suggest that this soft component arises from cooling of the forward shock similar to that found in Type IIn SN2010jl.

**Keywords:** Supernovae; stars:mass-loss; X-rays

(Received 1 September 2023; revised 30 April 2024; accepted 26 May 2024)

## 1. Introduction

Massive stars ($\gtrsim 8 \, M_\odot$) end their lives as core-collapse supernovae (SNe) once degeneracy pressure support is overcome by electron capture and nuclear photodissociation, leading to runaway collapse (e.g. Burbidge et al. 1957; Iben 1974; Woosley et al. 2002; Eldridge & Tout 2004; Smartt 2009; Janka 2012; Ibeling & Heger 2013; Burrows 2013). However, the nature of the explosion is thought to depend on the properties of the progenitor (e.g. De Donder & Vanbeveren 1998; Woosley et al. 2002; Heger et al. 2003; Izzard, Ramirez-Ruiz, & Tout 2004; Zapartas et al. 2017), as well as the type and strength of mass loss before core collapse (Smith 2014). There have been numerous attempts to identify SN progenitor stars in archival space- and ground-based images (e.g. Smartt et al. 2015). The majority of progenitors discovered in pre-explosion images are red supergiants (RSGs) with masses $\lesssim 20 \, M_\odot$ which are associated with hydrogen-rich Type II SNe (Smartt 2009; Maund et al. 2013), although there have also been progenitor stars associated with hydrogen-poor Type IIb SNe (Aldering, Humphreys, & Richmond 1994; Crockett et al. 2008; Maund et al. 2011; Van Dyk et al. 2014) and hydrogen-stripped Type Ib SNe and hydrogen-stripped Type Ib SNe (Cao et al. 2013; Eldridge et al. 2015; Kilpatrick et al. 2021).

Prior to core collapse, it is expected that RSGs will lose mass via winds. However, there is evidence that these winds are not strong enough to explain the properties of SNe that exhibit strong interaction with dense material (e.g. Beasor et al. 2020). While historically, massive star evolution was thought to be dominated by single-star wind mass loss, there is now strong evidence that $\sim 10\%$ of massive stars undergo substantial and even eruptive mass loss a few years prior to core collapse (see e.g. Smith 2014, and references therein). This can result in up to 1 $M_\odot$ of material being ejected in the decades to years before explosion (e.g. Smith 2014; Tinyanont et al. 2019; Margutti et al. 2014; Mauerhan et al. 2013; Kilpatrick et al. 2021), creating a dense shell of circumstellar material (CSM) around the progenitor with which the SN shock and ejecta interact (e.g. Smith et al. 2017). While the mechanism for such mass loss in single stars is typically attributed to winds or violent eruptions (e.g. Matsumoto & Metzger 2022), the frequency and cause of these extreme mass ejections remain uncertain. However, nuclear burning instabilities or interaction with a binary companion have been suggested as possible explanations (e.g. Quataert & Shiode 2012;

Smith 2014; Smith & Arnett 2014; Kochanek 2019; Sun et al. 2020; Matsuoka & Sawada 2023).

To shed light on pre-explosion behaviour, numerous studies have attempted to determine the frequency of these outbursts by searching for evidence of pre-SN variability in the form of precursor outbursts. These studies include the detection of precursor emission in individual supernovae such as SN2009ip (Fraser et al. 2013; Margutti 2014; Smith et al. 2022), SN2010mc (Ofek et al. 2013), SN2015bh (Elias-Rosa et al. 2016; Jencson et al. 2022), SN2016bhu (Pastorello et al. 2018), SN2020ltf (Jacobson-Galán et al. 2022a), and others (e.g. Ofek et al. 2014, 2016; Tartaglia et al. 2016; Kilpatrick et al. 2018; Strotjohann et al. 2021), and long-term, deep, high-resolution imaging surveys such as that completed with the Large Binocular Telescope (e.g. Kochanek et al. 2008; Johnson, Kochanek, & Adams 2017, 2018; Neustadt, Kochanek, & Smith 2023; Rizzo Smith, Kochanek, & Neustadt 2023). Johnson et al. (2018) found that the progenitors of a sample of Type II-P/L SNe exhibited up to 10% variability in the decade prior to explosion, with no more than 37% of these exhibiting outburst prior to explosion. Strotjohann et al. (2021) suggested that ∼25% of Type IIn SNe experienced precursor outbursts in the three months prior to explosion with luminosities >5 × 10^40 erg s^−1, the detection of which can be used to constrain the mass-loss rate and mechanism (Matsumoto & Metzger 2022).

Apart from looking for precursor emission, core-collapse SNe exhibit photometric and spectroscopic evidence of enhanced mass loss (Smith et al. 2011; Shivvers et al. 2015). For example, the presence of dense CSM from end-of-life mass loss can appear as narrow emission lines in the early (∼ days to hours after shock breakout) optical SN spectra (e.g. Gal-Yam et al. 2014; Kiewe et al. 2012; Jacobson-Galán et al. 2022a; Terreran et al. 2021; Tinyanont et al. 2022). These emission lines, commonly referred to as 'flash' ionisation features, allow the composition and density of the CSM to be probed, and thus proide insight into the progenitor and its mass-loss history up to radii of ∼10^15 cm (e.g. Khazov et al. 2016; Yaron et al. 2017; Bruch et al. 2021; Jacobson-Galán et al. 2022a; Boian & Groh 2020). For narrow emission lines from CSM interaction to be observed and not buried within the signal from the SN photosphere, mass-loss rates of >10^−2–10^−3 M_⊙ are generally required (Smith 2014; Fransson & Jerkstrand 2015).

The shock heating of the CSM not only provides insight into the mass-loss history of a progenitor during its final stages of evolution but also produces significant X-ray, UV, and radio emission (Chevalier & Fransson 2006). As the X-ray emission depends on the CSM density and the evolutionary parameters of the SN and its progenitor, one can then use the X-ray properties to constrain the wind density and progenitor's mass-loss history (e.g. Chevalier 1982; Chevalier & Fransson 2006; Dwarkadas & Gruszko 2012; Chandra 2018).

X-rays have been detected from a growing number of interacting transients (see Smith 2017; Dwarkadas & Gruszko 2012 and Chandra 2018 for reviews) including the Type IIn SNe 2005kl, 2006jd, and 2010jl (Chandra et al. 2015; Katsuda et al. 2016, e.g.), the interacting Type Ib SN2014C (Margutti et al. 2017; Brethauer et al. 2022), and the peculiar fast, blue optical transient AT2018cow (Xiang et al. 2021; Margutti et al. 2018a; Kuin et al. 2019; Rivera Sandoval & Maccarone 2018; Savchenko et al. 2018; Miller et al. 2018). Since these X-rays are indicative of interaction with dense material surrounding the progenitor, their detection can help to reveal the precise structure of the CSM. For example, the onset of X-rays from SN2014c after ∼20 days indicated the presence of a low-density cavity surrounding the progenitor which extended out to a radius of $R \sim (0.8 - 2) \times 10^{16}$ cm. From this, Margutti et al. (2017) inferred that significant mass loss did not occur within the last 7 yr of the progenitor's life, assuming a wind velocity of 1 000 km s^−1. In addition, Katsuda et al. (2016) suggested an aspherical, torus-shaped CSM for a sample of Type IIn SNe based on the spectral evolution of the X-ray emission from a single-component, heavily absorbed model to a moderately and heavily absorbed two-component, thermal model.

In this paper, we present a comprehensive study of both the multi-wavelength emission observed prior to the explosion of SN2023ixf, and the X-ray emission detected after first light by the *Neil Gehrels Swift Gamma-ray Burst Mission* (Gehrels et al. 2004, hereafter *Swift*). This builds on previous studies that have examined the infrared (e.g. Szalai & Dyk 2023; Kilpatrick et al. 2023; Jencson et al. 2023) and optical variability (e.g. Neustadt et al. 2023; Dong et al. 2023), and complements those that have focused on the higher energy post-explosion X-ray emission (e.g. Grefenstette et al. 2023) by making use of soft X-ray observations from *Swift*. In Section 2, we summarise the current knowledge of SN2023ixf at the time of writing. In Section 3, we present our observations, while in Section 4, we present our analysis of the pre-explosion properties and post-explosion X-rays. In Section 5, we present our discussion before ending with our summary and conclusions in Section 6.

## 2. SN2023ixf

In this section, we provide an overview of the properties of SN2023ixf and its progenitor based on the current literature and available data.

SN2023ixf was discovered on 19 May 2023 17:27 UTC (Itagaki 2023) and is one of the closest and brightest core-collapse SNe of the last decade, reaching peak absolute *u*- and *g*-band magnitudes of −18.6 and −18.4, respectively (Jacobson-Galan et al. 2023). Located in the host galaxy M101 (RA = 14:03:38.580, DEC = +54:18:42.1) at a distance of 6.4±0.2 Mpc (Shappee & Stanek 2011) and at a redshift of $z = 0.000804$ (Perley et al. 2023), it presents a valuable opportunity to study the evolution of a core-collapse SN in detail.

SN2023ixf's evolution has been extensively monitored, with numerous facilities, and professional and amateur astronomers have reported early photometric (e.g. Villafane et al. 2023; Kendurkar & Balam 2023a,b; Brothers et al. 2023; Pessev et al. 2023; Vannini 2023; Desrosiers, Kendurkar, & Balam 2023; Fowler, Sienkiewicz, & Dussault 2023; D'Avanzo et al. 2023; Silva 2023a; Vannini & Julio 2023a,b; Balam & Kendurkar 2023; Maund et al. 2023; Singh et al. 2023; Koltenbah 2023; Mayya et al. 2023; Kendurkar & Balam 2023c; Chen et al. 2023; Silva 2023b; Daglas 2023; Sgro et al. 2023) and spectroscopic (e.g. Sutaria & Ray 2023; Sutaria, Mathure, & Ray 2023; Zhang et al. 2023b; González-Carballo et al. 2023; Stritzinger et al. 2023; BenZvi et al. 2023a,b; Lundquist, O'Meara, & Walawender 2023) observations.

Analysis of pre-discovery data from the Zwicky Transient Facility (ZTF) (Perley & Irani 2023), Asteroid Terrestrial-impact Last Alert System (ATLAS) (Fulton et al. 2023), and many other telescopes constrained the explosion time to between 19:30 and 20:30 UTC on 18 May 2023 (Yaron et al. 2023; Chufarin et al. 2023; Zhang et al. 2023a; Limeburner 2023; Mao et al. 2023; Hamann 2023; Filippenko, Zheng, & Yang 2023). In our analysis and similar to Jacobson-Galan et al. (2023), we adopt the time of first light as MJD 60082.83 ± 0.02 (Mao et al. 2023).

Shortly following its detection, SN2023ixf was classified as a Type II SN based on the strong flash ionisation features of H, He, C, and N in its spectrum, as well as the subsequent emergence of broad P-cygni features from H and He (Perley et al. 2023; Perley & Gal-Yam 2023; Perley 2023; Jacobson-Galan et al. 2023; Teja et al. 2023; Hiramatsu et al. 2023a; Bianciardi et al. 2023). In addition, a strong blue continuum with Balmer emission and features of He II, N IV, and C IV, as well as similarities to other Type II SNe including SN2014G, SN2017ahn, and SN2020pni (Bostroem et al. 2023), led Yamanaka, Fujii, & Nagayama (2023) to suggest that SN2023ixf is a high-luminosity Type II SN embedded in a nitrogen- and helium-rich CSM. Optical spectropolarimetry of SN2023ixf revealed a high-continuum polarisation of ~1% up to day 2.5 post-explosion, which decreased to ~0.5% on day 3.5 and persisted until day 14.5. This decline coincided with the disappearance of the flash ionisation features from its spectrum (Vasylyev et al. 2023). Vasylyev et al. (2023) attributed the temporal evolution of the polarisation to an aspherical explosion in a highly-asymmetric CSM that was carved out by pre-explosion mass loss in the progenitor.

Comparisons to light curve and spectral models from the non-LTE radiative transfer code CMFGEN and the radiation hydrodynamics code HERACLES, along with the disappearance of the narrow emission lines from SN2023ixf's spectrum a few days after the explosion (Smith et al. 2023), indicate that the progenitor had a dense, solar-metallicity CSM that extended out to a radius of $(0.5-1) \times 10^{15}$ cm (Jacobson-Galan et al. 2023), which is consistent with a wind with a density that decreases following $r^{-2}$ (Kochanek 2019). Jacobson-Galan et al. (2023) also determined an enhanced progenitor mass-loss rate of $10^{-2} \, M_\odot \, yr^{-1}$ in the 3–6 yr prior to explosion, with Bostroem et al. (2023) finding a value of $10^{-3}$–$10^{-2} \, M_\odot \, yr^{-1}$ when comparing to additional CMFGEN models. This is in contrast to Jencson et al. (2023), who estimated a lower pre-explosion mass-loss rate between $3 \times 10^{-5}$ and $3 \times 10^{-4} \, M_\odot \, yr^{-1}$ in the final 3–19 yr before explosion using near- and mid-infrared (IR) spectral energy distribution (SED) modelling, while the SED fits to *Spitzer* and *HST* data by Niu et al. (2023) yielded a value of $1 \times 10^{-5} \, M_\odot \, yr^{-1}$.

In addition, Jencson et al. (2023) found no evidence of pre-explosion outbursts in *Spitzer* data, and instead favoured a scenario where a steady, enhanced ejected material for >13 yr out to a radius of $>4 \times 10^{14}$ cm. This is consistent with the results of Neustadt et al. (2023), who found no evidence of outbursts in the progenitor's final 15 yr and a mass-loss rate of $\sim 10^{-5} \, M_\odot \, yr^{-1}$. Using the empirical mass-loss rate prescription from Goldman et al. (2017), Soraisam et al. (2023c) estimated a mass-loss rate of $(2 - 4) \times 10^{-4} \, M_\odot \, yr^{-1}$, in closer agreement with the results of Jencson et al. (2023) and Neustadt et al. (2023) than those from Jacobson-Galan et al. (2023) and Bostroem et al. (2023). The current mass-loss rate estimates for SN2023ixf are summarised in Fig. 1, and the discrepancies between these values are examined in Section 5.2.2.

From the emergence of a multi-peaked emission line profile of H$\alpha$ at $t \sim 16$ days, Teja et al. (2023) proposed a shell-shaped CSM with inner and outer radii of $\sim 8.5 \times 10^{14}$ cm and $\sim 20.9 \times 10^{14}$ cm, respectively, corresponding to enhanced mass loss $\sim 35$–65 yr before the explosion. The flash ionisation features and U–V colour of SN2023ixf led Hiramatsu et al. (2023a) to suggest a delayed shock breakout due to a dense CSM with a radial extent of $\sim (3$–7$) \times 10^{14}$ cm. Under the assumption of continuous mass loss, Hiramatsu et al. (2023a) determined a mass-loss rate of

0.1–1 $M_\odot \, yr^{-1}$ in the final 1–2 yr before explosion, with the rate decreasing to 0.01–0.1 $M_\odot \, yr^{-1}$ in the $\sim 0.4$–0.7 yr to explosion. Assuming an eruptive mass-loss mechanism, they proposed these eruptions released 0.3–1 $M_\odot$ roughly one year prior to explosion.

Pre-explosion *HST* and *Spitzer* images have since revealed a point source at the SN position that is consistent with a RSG progenitor shrouded by a large amount (0.1–1 $M_\odot$, Hiramatsu et al. 2023a) of CSM (Soraisam et al. 2023b,a; Mayya 2023; Jencson et al. 2023; Soraisam et al. 2023c; Neustadt et al. 2023). However, no UV nor X-ray counterpart can be seen in *AstroSat-UVIT* (Basu et al. 2023), *Chandra* (Kong 2023), or *XMM-Newton* images (Matsunaga et al. 2023), consistent with our study.

From the pre-explosion data, progenitor masses of $11 \pm 2 \, M_\odot$ (Kilpatrick et al. 2023), $\sim 12 \, M_\odot$ (Pledger & Shara 2023), $\sim 15 \, M_\odot$ (Szalai & Dyk 2023), $17 \pm 4 \, M_\odot$ (Jencson et al. 2023), $20 \pm 4 \, M_\odot$ (Soraisam et al. 2023c), 16.2–17.4 $M_\odot$ (Niu et al. 2023), and $\sim 9.3$–13.6 $M_\odot$ (Neustadt et al. 2023) have been suggested. In addition, the star formation histories constructed by Niu et al. (2023) using the surrounding resolved stellar populations indicated a higher-mass, Type II SN progenitor in the 17–19 $M_\odot$ range. Similarly, shock cooling emission models of the light curve indicated a progenitor radius of $410 \pm 10 \, R_\odot$, consistent with a RSG (Hosseinzadeh et al. 2023), and in agreement with RSG designations based on HR evolutionary tracks (Kilpatrick et al. 2023; Jencson et al. 2023), the progenitor's IR colour and pulsations (Soraisam et al. 2023c), and SED modelling of the progenitor (Neustadt et al. 2023).

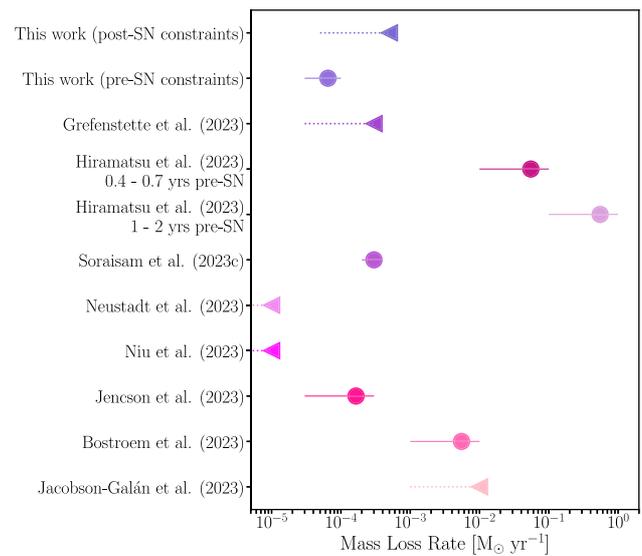

**Figure 1.** A summary of the mass-loss rates derived from both this study and other studies presented in the literature. Time scales (where available) and methods relevant to each analysis: this work (post-SN): formalism from Margutti et al. (2012) and Margutti et al. (2018b), >0.5–1.5 yr prior to explosion; this work (pre-SN): mass-loss prescription from Matsumoto & Metzger (2022), Grefenstette et al. (2023): *NuSTAR* post-explosion ($t < 11$ days); Hiramatsu et al. (2023a): numerical light-curve modelling; Soraisam et al. (2023c): mass-loss rate prescription from Goldman et al. (2017), Neustadt et al. (2023): mass-loss prescription from Matsumoto & Metzger (2022), Niu et al. (2023): SED modelling and mass-loss prescription from Beasor & Davies (2016), Jencson et al. (2023): SED modelling, 3–19 yr prior to explosion; Bostroem et al. (2023): CMFGEN spectral modelling; Jacobson-Galán et al. (2023): light curve and spectral modelling, ~3–6 yr prior to explosion.

Szalai & Dyk (2023) identified no significant flux variability in pre-explosion *Spitzer* images collected between 2004 and 2019 that might indicate eruptive mass-loss activity. Likewise, no variability was identified during the progenitor's final 15 yr using optical pre-explosion data from the Large Binocular Telescope (Neustadt et al. 2023), or in pre-explosion ATLAS, ZTF and WISE data (Hiramatsu et al. 2023a). In addition, UV observations from *GALEX* showed no variability in the 15–20 yr prior to explosion (Flinner et al. 2023). Performing a more detailed analysis of the *Spitzer* data, Kilpatrick et al. (2023) identified a 2.8 yr timescale variability in *Spitzer* 3.6 and 4.6 $\mu$m imaging, which was subsequently confirmed by Jencson et al. (2023) and Soraisam et al. (2023c). These authors attributed this variability to pulsations as opposed to an eruptive outburst.

Searches for pre-explosion optical outbursts in Distance Less Than 40 MPc survey (DLT40), ATLAS, and ZTF data by Dong et al. (2023) found a low probability of significant outbursts in the five years prior to the explosion. They also inferred a maximum ejected CSM mass of $\sim 0.015$ $M_{\odot}$, leading them to suggest that if the dense CSM surrounding SN2023ixf is the result of one or more precursor outbursts, they were likely faint, of short duration ($\sim$ days to months), or occurred more than 5 yr before explosion. Based on their analysis, Dong et al. (2023) proposed that more than one physical mechanism may be responsible for the dense CSM observed in SN2023ixf, such as the interaction of stellar winds from binary companions (e.g. Kochanek 2019).

No emission was initially detected at radio frequencies (Berger et al. 2023b; Matthews et al. 2023b; Chandra et al. 2023a; Matthews et al. 2023c); however, statistically significant emission with a flux density of $41 \pm 8$ $\mu$Jy was detected 29 days after discovery (Matthews et al. 2023a). Using 1.3 mm (230 GHz) observations, Berger et al. (2023a) determined upper limits of $8.6 \times 10^{25}$ erg s$^{-1}$ at 2.7 and 7.7 days, and $3.4 \times 10^{25}$ erg s$^{-1}$ at 18.6 days. Searches for neutrinos (Thwaites et al. 2023; Nakahata & Super-Kamiokande Collaboration 2023; Guetta et al. 2023) and gamma rays (Marti-Devesa 2023) from SN2023ixf are consistent with background. Using the derived upper limits from the *Fermi-LAT* gamma-ray flux and the IceCube neutrino flux, in addition to the shock and CSM properties, Sarmah (2023) estimated of $\sim 10^{-11}$ erg cm$^2$ s$^{-1}$ on the gamma-ray flux and $\sim 10^{-3}$ GeV cm$^{-2}$ on the neutrino fluence for emission produced via the proton-proton chain.

However, X-ray data reveal a point source at the location of SN2023ixf post explosion. *NuSTAR* observations in the 3–20 keV range show a highly absorbed continuum with a strong emission line at $\sim$6.4 keV, likely attributable to Fe. The extrapolated broadband flux in the 0.3–30 keV range of $2.3 \times 10^{-12}$ erg cm$^2$ s$^{-1}$ yields an intrinsic, absorbed X-ray luminosity of $1.1 \times 10^{40}$ erg s$^{-1}$ (Grefenstette 2023). A subsequent detailed analysis of *NuSTAR* observations yielded an absorbed X-ray luminosity of $2.5 \times 10^{40}$ erg s$^{-1}$ in the 0.3–79 keV range, assuming a hot, thermal-bremsstrahlung continuum ($T > 25$ keV), from which a progenitor mass-loss rate of $\dot{M} \sim 3 \times 10^{-4}$ $M_{\odot}$ yr$^{-1}$ was calculated (Grefenstette et al. 2023).

The best-fit model to *Chandra* observations was found to be a $\sim$10 keV plasma with a column density of $\sim 3.2 \times 10^{22}$ cm$^{-2}$, consistent with a normal stellar wind (Kochanek 2019). From this, Chandra et al. (2023b) estimated an unabsorbed flux in the 0.3–10 keV band of $(1.6 \pm 0.1) \times 10^{-12}$ erg cm$^2$ s$^{-1}$, corresponding to a luminosity of $8 \times 10^{39}$ erg s$^{-1}$ at a distance of 6.4 Mpc.

Hard X-rays from *ART-XC* onboard the *SRG* Observatory were instead found to be best described by an absorbed power-law model with a narrow Gaussian at 6.4 keV and a column density fixed to $2 \times 10^{23}$ cm$^{-2}$. In addition, the *ART-XC* data showed no variability across a timescale of $\sim$10 ks in the 4–12 keV band (Mereminskiy et al. 2023).

Follow-up *Swift* target-of-opportunity observations identified $8.8 \pm 3.7$ background-subtracted counts in a 20-arcsec region around SN2023ixf's reported position. Assuming an absorbed power law with an index of 1.3 and a column density of $2 \times 10^{23}$ cm$^{-2}$, Kong (2023) found an absorbed flux in the 0.03–30 keV range of $7.7 \times 10^{-13}$ erg cm$^2$ s$^{-1}$, corresponding to a luminosity of $3.8 \times 10^{39}$ erg s$^{-1}$ at 6.4 Mpc. Additionally, Grefenstette et al. (2023) determined a 0.3–10 keV flux of $6.6^{+10}_{-6.6} \times 10^{-14}$ erg cm$^2$ s$^{-1}$ using a stacked spectrum consisting of 25 observations spanning 6 days post explosion, a value $\sim$9 times lower than their result using *NuSTAR* observations.

In summary, SN2023ixf is a nearby, interacting, low-luminosity Type II SN from a RSG progenitor that shows signatures of enhanced mass loss during its final years.

## 3. Observations

SN2023ixf and its progenitor have been extensively observed at multiple wavelengths. In this section, we provide an overview of the UV, optical and X-ray facilities relevant to this study and describe our data reduction and preparation techniques.

### 3.1 *Optical/UV photometry*

In Figure 2, we show the pre-explosion UV/optical light curves of SN2023ixf from the All-Sky Automated Survey for Supernovae (ASAS-SN), the Zwicky Transient Facility (ZTF), the Asteroid Terrestrial-impact Last Alert System (ATLAS), and *Swift*. In Figure 3, we show the post-explosion optical/UV light curves of SN2023ixf.

#### 3.1.1 *ASAS-SN*

ASAS-SN (Shappee et al. 2014; Kochanek et al. 2017; Hart et al. 2023) is an automated optical transient survey that monitors the visible sky every $\sim$20 hours to a depth of $\sim$18.5 mag in the *g*-band. Starting in late 2011, ASAS-SN began observing the northern sky in the *V*-band which had a depth of $\sim$17.5 mag. In 2017–2018, the survey switched to the *g*-band to gain a magnitude of depth and added additional units for cadence. The survey now consists of 20 individual telescopes with 14-cm aperture Nikon telephoto lenses with $\sim$8 arcsec pixels that are grouped together into five 4-telescope units. The five telescope units are located at Haleakala Observatory, McDonald Observatory, the South African Astrophysical Observatory, with the remaining two at the Cerro Tololo Inter-American Observatory.

We used the Sky Patrol V1[a] (Kochanek et al. 2017) to to obtain image-subtracted light curves for the location of SN2023ixf in addition to a grid of 13 points adjacent to the SN location. The Sky Patrol uses images reduced by the ASAS-SN fully automated pipeline which includes the ISIS image subtraction package (Alard & Lupton 1998; Alard 2000). When serving image-subtracted light curves, the Sky Patrol first co-adds the subtracted images from the 3 dithers taken at each pointing during survey operations. To derive the *g*- and *V*-band photometry from these images, Sky Patrol V1 uses the IRAF apphot package with a 2-pixel (or ∼16 arcsec) radius aperture to perform aperture photometry on each subtracted image, generating a differential light curve. The photometry is calibrated using the AAVSO Photometric All-Sky Survey (Henden et al. 2015). Images with a full width at half maximum of 1.7 pixels or greater and images with a shallow depth (3$\sigma$ detection limit of <18.4 mag) were discarded. ASAS-SN observed M101 665 times from Jan 2012 to Nov 2018 in the *V*-band and 864 times from Nov 2017 through May 2023 in the *g*-band.

----

[a] https://asas-sn.osu.edu/.

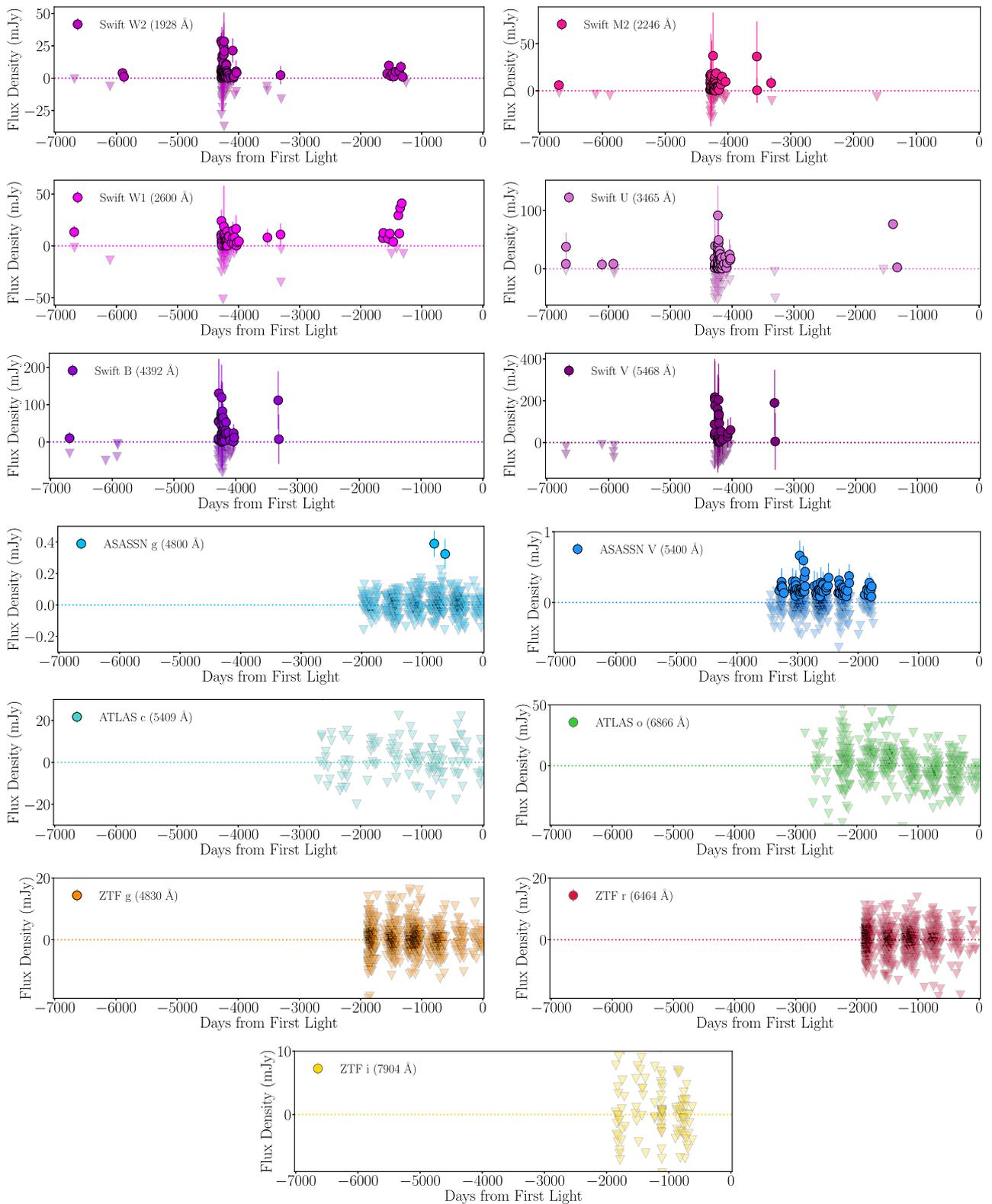

**Figure 2.** Pre-explosion UV/optical light curves of SN2023ixf as seen by *Swift*, ASAS-SN, ZTF, and ATLAS. Here, solid data points correspond to fluxes that are $\geq 3\sigma$ above the reference flux in that band, while the shaded triangles indicate that the emission is consistent with the reference flux in that band.

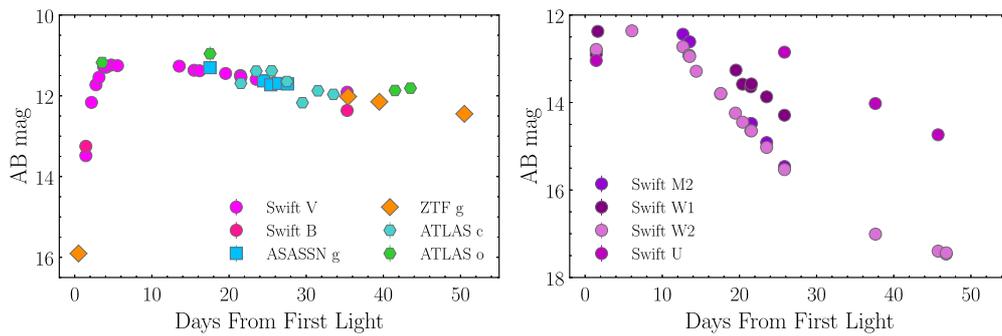

**Figure 3.** Post-explosion optical (left panel) and UV (right panel) light curves of SN2023ixf as seen by *Swift*, ASAS-SN, ATLAS, and ZTF. Here, only data detected with $\geq 3\sigma$ detection significance is shown.

### 3.1.2 ATLAS

ATLAS currently consists of a quadruple 0.5-m telescope system that has two units in Hawaii (Haleakala and Mauna Loa), one in Chile (El Sauce), and one in South Africa (Sutherland). The ATLAS observing strategy is to obtain a sequence of four 30-s exposures spread out over an hour (Smith et al. 2020), using either a cyan (*c*) filter [4,420–650 nm] or an orange (*o*) filter [560–820 nm], depending on the Moon phase (Tonry et al. 2018a). The pixel scale of ATLAS is ∼1.9 arcsec/pixel and the typical PSF is ∼4 arcsec.

ATLAS data are accessible through the ATLAS Forced Photometry Server[b] (Shingles et al. 2021). The photometric routine used, called tphot, is based on algorithms described in Tonry (2011) and Sonnett et al. (2013) and can be deployed on either reduced or difference images. Both types of images have been calibrated astrometrically and photometrically using the ATLAS All-Sky Stellar Reference Catalog (Refcat, Tonry et al. 2018b), and the difference images also use a modified version of the image subtraction algorithm HOTPANTS (Becker 2015) to subtract a reference sky frame.

Counting each 30-s exposure individually, ATLAS has observed the location of SN2023ixf 2491 times between 30 July 2015 and 2 July 2023. We do not use photometry from images where the location of SN2023ixf is within 40 pixels of a chip edge (10 images), where the best-fit axis ratio of a source at that location is above 1.5 (16 images), where the 5$\sigma$ limiting magnitude is less than 16 (56 images), or where the ATLAS pipeline presents an error flag (no images). This leaves us with 1 844 *o*-band and 548 *c*-band images.

We then co-added nightly observations, discarding observations with flux values over three times their uncertainty from the median flux, weighting the remaining observations by their inverse variance in flux, and recording the interpolated 50th percentile values for epoch, flux, and uncertainty. The final light curve has 438 *o*-band entries and 123 *c*-band entries.

### 3.1.3 Zwicky transient facility

ZTF *g*-, *r*-, and *i*-band photometry covering the full ZTF survey from 17 March 2018 (JD = 2 458 194.5) until 08 July 2023 (JD = 2460133.8) were obtained using the ZTF forced photometry service (Masci et al. 2019). This covers a phase of ∼1 850 days prior to discovery until ∼50 days after discovery of SN2023ixf. Following the procedure outlined in the ZTF forced photometry manual v2.3,[c] we apply the standard baseline corrections to the data and use a signal-to-noise threshold of 3 for all available data. We note that ZTF has a characteristic pixel scale of ∼ 1 arcsec/pixel and a PSF of ∼ 2 arcsec.

### 3.1.4 *Swift* UVOT

Due to its proximity and the discovery of SN2011fe in 2011, M101 has been extensively monitored by *Swift*. Prior to the discovery of SN2023ixf, there had been 218 observations overlapping the location of this event. These observations were carried out between 29 August 2006 (*MJD* = 53976.48613) and 08 December 2019 (MJD=58825.72311) and have *Swift* target IDs of 35892, 30896, 32081, 32088, 32094, 32101, 33635, 11002, and 32481. These observations were conducted using both the UltraViolet and Optical Telescope (Roming et al. 2005, UVOT) and the X-ray Telescope (Burrows et al. 2005, XRT). The total exposure time of these observations is ∼431 ks.

Since the discovery of SN2023ixf until September 2023, *Swift* has observed the source 58 times. These observations began on 20 May 2023 (*MJD* = 60 084.26901) and have *Swift* target IDs of 16038, 16043, 32481, and 89625. The total combined exposure of these observations is ∼68 ks. For the majority of *Swift* observations, *Swift* observed either SN2023ixf or the location of the SN using at least one or more of the six UVOT filters (Poole et al. 2008:[c] *V* (5 425.3 Å), *B* (4 349.6 Å), *U* (3 467.1 Å), *UVW*1 (2 580.8 Å), *UVM*2 (2 246.4 Å), and *UVW*2 (2 054.6 Å)).

To derive the UVOT photometry both prior to and during the rise of SN2023ixf, we used the HEASARC UVOTSOURCE package.

---

[b]https://fallingstar-data.com/forcedphot/.

[c]https://irsa.ipac.caltech.edu/data/ZTF/docs/ztf_forced_photometry.pdf.

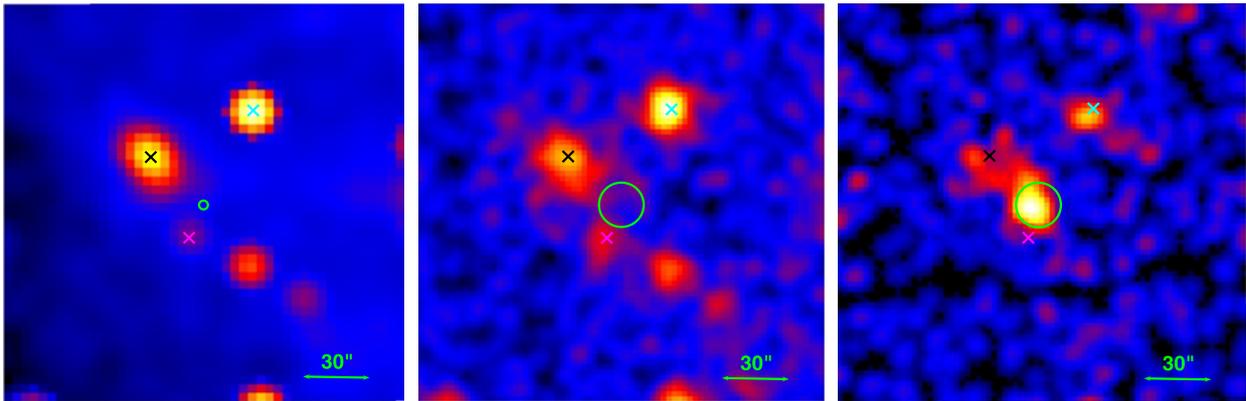

**Figure 4.** *(Left Panel):* The merged, broadband, pre-explosion *Chandra* observation of the location of SN2023ixf. The two arcsecond radius green circle shows the location of SN2023ixf, and the black, cyan, and magenta crosses mark the locations of the high-mass X-ray binaries (HMXBs) CXO J140341.1+541903, CXO J140336.1+541924, and [CHP2004] J140339.3+541827, respectively. *(Middle Panel):* The merged, broadband *Swift* observation obtained using all available pre-explosion observations. The green circle here has a radius of 15 arcsec and is centered on SN2023ixf used to derive count rates. *(Right panel:)* The merged, broadband *Swift* observation created using all available post-explosion observations. Here the green circle has a 20 arcsec radius and is centered on SN2023ixf. Significant X-ray emission arises from the location of the source. Note that the images are all aligned to a common reference frame.

To extract the UV and optical counts, we used a circular region with a 5-arcsec radius centered on the position of SN2023ixf and source-free background regions with a radius of 20 arcsec located at $(\alpha, \delta) = (14{:}03{:}42.5088, +54{:}18{:}12.172)$ and $(\alpha, \delta) = (14{:}03{:}31.6839, +54{:}16{:}03.572)$ for the pre- and post-discovery data, respectively. Similar to Hosseinzadeh et al. (2023), we find that the majority of the early post-explosion *Swift* UVOT observations were saturated due to the high count rate of the source and thus we discard these data and only use those that are below the maximum countrate limit of the UVOT detector. The UVOT count rates are converted into AB magnitudes and flux densities using the most recent calibrations (Poole et al. 2008; Breeveld et al. 2010). We do not correct the photometry for Galactic extinction.

### 3.2 X-ray observations

#### 3.2.1 *Swift* XRT

In addition to the UVOT, *Swift* simultaneously observed the location and rise of SN2023ixf using the XRT in photon-counting mode. Following the *Swift* XRT reduction guide, we reduced all observations using the standard filter and screening criteria, and the most up-to-date calibration files. Using the task *XRTPIPELINE*, we reprocessed all level one XRT data, producing cleaned event files and exposure maps for all observations.

To increase the signal-to-noise ratio of our observations, we combined our individual images using *XSELECT* version 2.5b. Here, we combined the observations taken prior to the discovery of SN2023ixf into 33 time bins, spanning ∼6 000 days since the first serendipitous *Swift* observation of this location. For the observations taken post-discovery, we combined these observations into 19 time bins spanning ∼46 days after first light. We also merged all observations taken prior to and after the discovery of SN2023ixf to produce a deep *Swift* XRT observation for both before and after the discovery of SN2023ixf (see Fig. 4 middle and right panel).

Background-subtracted count rates were derived from each of these merged observations using a 20 arcsec source region centered on the position of SN2023ixf for observations taken after discovery and a 15-arcsec source region for observations taken prior to discovery. A smaller source region was used for the pre-explosion observations due to the presence of an X-ray bright source near the location of SN2023ixf in both the deep *Chandra* and *Swift* images that were obtained by merging all available observations taken prior to peak (see Fig. 4 left and middle panel). However, as these sources were only observable in these deep exposures and were not seen in the individual observations taken post-explosion due to the brightness of SN2023ixf, we use a larger region to maximise the flux from this event. The total exposure time of these merged observations was 431.2 and 68 ks, for the pre- and post-explosion observations, respectively. For all observations we used the same source-free background region with a radius of 100 arcsec centered at $(\alpha, \delta) = (14{:}03{:}27.7541, +54{:}14{:}16.237)$. This region, as well as the background regions used for the Chandra and XMM-Newton observations, were selected to capture the local diffuse emission and background contribution. All extracted count rates were corrected for the encircled energy fraction (Moretti et al. 2004).

Using our merged event files, we extracted a spectrum from the position of the source using the *Swift* analysis tool *XRTPRODUCTS* and the source and background regions defined above. Ancillary response files were generated using the task *XRTMKARF* and the individual exposure maps from *xrtpipeline* that were merged using the HEASARC analysis tool *ximage* version 4.5.1. The response matrix files were obtained from the most recent calibration database. The resulting spectrum was grouped to have a minimum of 15 counts per energy bin using the FTOOLS command *grppha*.

#### 3.2.2 Chandra

The location of SN2023ixf has been serendipitously observed using the *Chandra X-ray Observatory* 21 times since 2000, with the most recent archival observation taken in 2017. These observations were taken in VFAINT or FAINT mode under the observation IDs: 934, 6175, 6170, 6169, 6152, 6118, 6115, 6114, 5323, 5322, 5309, 5300, 5297, 5296, 4737, 4736, 4735, 4733, 4732, 4731, 2065,

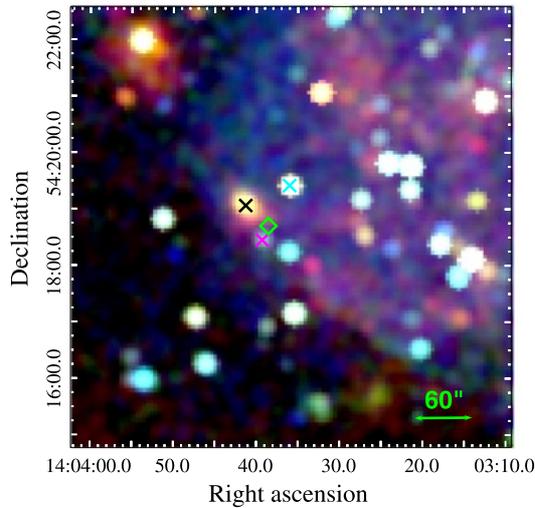

**Figure 5.** A merged and exposure-corrected *Chandra* X-ray image of the location of SN2023ixf (green diamond). The black, cyan, and magenta crosses mark the locations of HMXBs CXO J140341.1+541903, CXO J140336.1+541924, and [CHP2004] J140339.3+541827, respectively. Here, the 0.5–1.2 keV (soft) emission is in red, the 1.2–7.0 keV (medium) emission is in green, and the 2.0–7.0 keV (hard) emission is in blue.

and 19304. These observations were conducted using the ACIS-S detector and have a total combined exposure time of ∼966.2 ks.

All *Chandra* data were reduced using version 4.15 of the *Chandra* analysis software *CIAO*. We reprocessed the level one data using the *CIAO* command *chandra_repro* and the most up-to-date calibration database. To improve the absolute astrometry of these observations, we used the *CIAO* tool *wcs_match* and cross-matched X-ray sources found within these observations using the tool *wavdetect*, with the USNO-A2.0 catalog.[d] We then used *reproject_obs* to reproject these event files to a common tangent point based on the updated world coordinate system (WCS) information of the earliest and deepest *Chandra* observation in our dataset (ObsID: 934). This command also takes the reprojected event files and merges them together to form a single event file. We then used *flux_obs* to combine the reprojected observations to create an exposure-corrected image in the broad (see Fig. 4 left panel), soft, medium, and hard energy bands (see Fig. 5).

To place constraints on pre-explosion X-ray emission, we used a circular region centered on SN2023ixf with a radius of 2 arcsec and a source-free, local, background region located at $(\alpha, \delta) = $ (14:03:27.4629, +54:17:34.929) with a radius of 20 arcsec. A region of this size encloses 95% of all source photons at 1.496 keV, and as such, the extracted count rates were corrected for encircled energy fraction.[e]

### 3.2.3 XMM-Newton

*XMM-Newton* observed the location of SN2023ixf four times prior to its explosion. The first observation was taken in 2002, with the most recent from 2018. These observations have ObsIDs 0104260101, 0164560701, 0212480201, and 0824450501, totaling ∼220 ks prior to filtering. To analyse these observations, we used the *XMM-Newton* Science System (SAS) version 20.0.0 and the most up-to-date calibration file. Before extracting count rates, we produced cleaned event files by removing time intervals which were contaminated by a high background, or those during which proton flares were identified when generating count rate histograms for energies between 10.0–12.0 keV. We used the standard screening criteria as suggested by the current SAS analysis threads and *XMM-Newton* Users Handbook. For the MOS detectors, we used single to quadruple patterned events (PATTERN ≤ 12), while for the PN detectors only single and double patterned events (PATTERN ≤ 4) were selected. The standard canned screening set of FLAGS for both the MOS (#XMMEA_EM) and PN (#XMMEA_EP) detector were also selected.

[d] https://cxc.cfa.harvard.edu/ciao/threads/reproject_aspect/index.html#usno.
[e] http://cxc.harvard.edu/proposer/POG/html/chap4.html.

Count rates were extracted from a circular region with a radius of 20 arcsec centered on the location of SN2023ixf and a source-free, local, background region of radius 90 arcsec centered at $(\alpha, \delta) = $ (14:03:40.1033, +54:21:13.639). As a region of this size encloses only ∼80% of all source photons, all extracted counts were corrected for this aperture. Before extracting our counts, we combined both the MOS and PN detectors for each observation using the SAS tools command *merge*.

## 4. Analysis

In this section, we infer the properties of the progenitor of SN2023ixf by searching for evidence of pre-explosion variability and mass loss using both pre- and post-explosion observations.

### 4.1 Properties of the explosion

To constrain the pre-explosion mass loss using post-explosion X-ray observations (see Section 5.2.2), we require an estimate of the bolometric luminosity. As such, we use the SUPERBOL pipeline (Nicholl 2018) to calculate the bolometric light curve using the *Swift* UVOT light curves extracted post-explosion and the *u/U*, *B*, *g*, *V*, *r/R*, *i/I*, and *z* photometry published in Figure 1 of Jacobson-Galan et al. (2023). To show all light curves across the same timescale, we used the *i*-band as the reference filter and interpolated/extrapolated each light curve using a polynomial of order four between MJD 60082 and MJD 60100. We also assume a host reddening to SN 2023ixf of $E(B - V) = 0.033$ mag from Jacobson-Galan et al. (2023) that was derived from NaI D line absorption from the optical spectra of this event. We then fit the resulting SED with a blackbody model to derive the luminosity, radius, and temperature. In Fig. 6, we present our derived bolometric luminosity for SN2023ixf, along with the bolometric luminosities of other Type IIn SNe and the pseudobolometric luminosity for SN2023ixf from Hiramatsu et al. (2023a).

Our analysis in Section 5.2.2 also requires an estimate of the ejecta mass. From the analytical light curve model of Hiramatsu et al. (2023b) that attributes the SN emission to shock interaction between the ejecta and the CSM, we use $v = \sqrt{\frac{2(5-\delta)(n-5)E_{SN}}{(3-\delta)(n-3)M_{SN}}}$ to estimate the ejecta mass. Here, $M_{SN}$ is the ejecta mass, $E_{SN}$ is the explosion energy, the exponents of the broken power law that describes the density of the unshocked SN ejecta are $\delta = 0$ and $n = 12$ (as is standard for RSG progenitors), and $v$ is the characteristic velocity of the ejecta which corresponds to the photospheric velocity at maximum light (see Equation 4 from Hiramatsu et al. 2023b). For $v$, we use the lower limit on the SN shock velocity from Jacobson-Galan et al. (2023), who determined a value of 8 500 km s⁻¹ using the blue edge of the Hα absorption profile. Assuming $E_{SN} = (0.5 - 2) \times 10^{51}$ erg s⁻¹, we get an ejecta mass of 0.9–3.6 M$_\odot$, with an energy of $E_{SN} = 10^{51}$ erg s$^{-1}$ corresponding to an ejecta mass of 1.8 M$_\odot$.

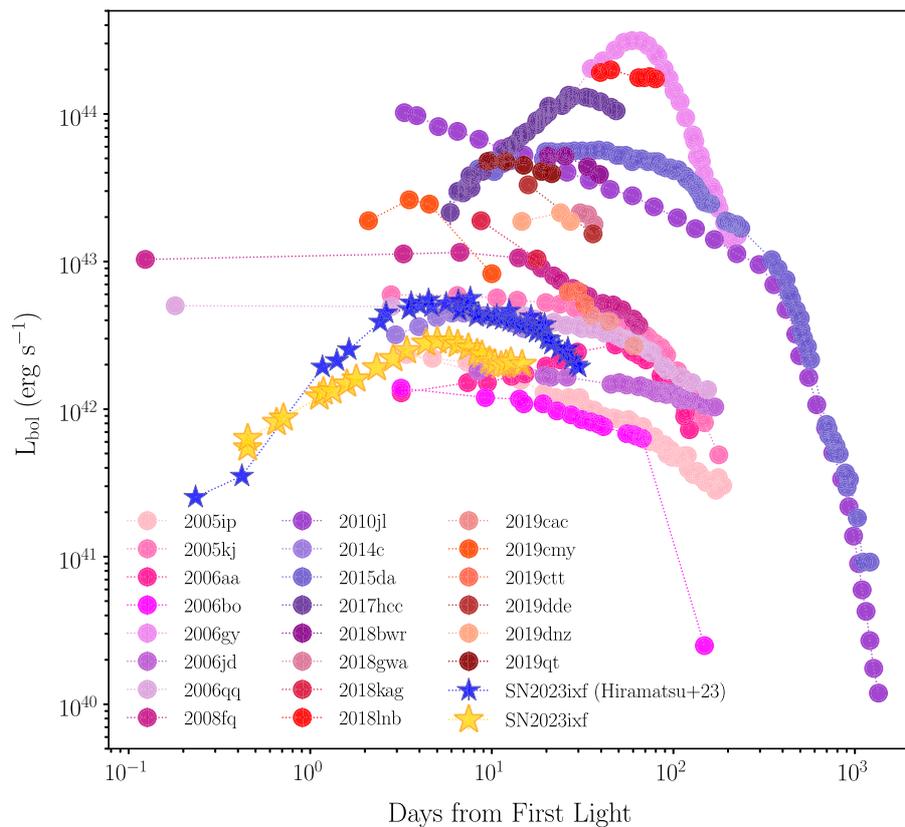

**Figure 6.** The bolometric light curve of SN2023ixf (yellow stars) compared to the pseudobolometric light curve (*UBVRI*, blue stars) from Hiramatsu et al. ([2023a](#)) and a sample of other Type IIn SNe. Data Sources: SN2006gy (Smith et al. [2010](#)); SN2010jl (Chandra et al. [2015](#)); SN2014c (Margutti et al. [2017](#)); SN2015da (Tartaglia et al. [2020](#)); SN2017hcc (Prieto et al. [2017](#)); SNe 2018bwr, 2018gwa, 2018kag, 2018lnb, 2019cac, 2019cmy, 2019ctt, 2019dde, 2019dnz, 2019qt (Soumagnac et al. [2020](#)); and all remaining SNe (Taddia et al. [2013](#)).

### 4.2 Variability

To constrain the presence of variability prior to explosion, we take advantage of the methods presented in Johnson et al. ([2017](#), [2018](#)) and Neustadt et al. ([2023](#)). Here, we examine the variability of the progenitor by calculating the peak-to-peak luminosity changes of the pre-SN differential light curves ($\Delta\lambda L_\lambda$). We then compare this variability to both the root mean square (RMS) of our data and to the RMS scatter of the comparison sample about the mean of their peak-to-peak luminosity changes. Our comparison sample consists of regions nearby the position of the SN from which these comparison light curves were extracted. This was done to better understand any systematic errors in the light curves. For our *Swift* and ZTF sample, we used a total of 12 control light curves, while for ATLAS and ASAS-SN we used 7 and 13, respectively. These sample points were chosen to avoid obvious nearby sources such as the optical/UV bright HMXBs shown in Figs. 4 and 5.

In Figs. 7–10, we present our resulting peak-to-peak luminosity variability analysis for both the progenitor of SN2023ixf and our control light curves.

Our plots suggest that for the majority of wavelengths we analysed, the observed scatter in the luminosity at the location of SN2023ixf is consistent with the comparison sample. This suggests that there is no evidence of pre-SN variability in these bands, consistent with the results of Neustadt et al. ([2023](#)) and Dong et al. ([2023](#)). We do see some evidence of possible dimming detected in the Swift W2 band nearly 6 000 days prior to explosion. However, when one considers the uncertainties in our analysis, these data points lie no more than $1.5\sigma$ from the mean of our controls, suggesting that it may not be significant. Similarly, for the ATLAS *o*-band, ASAS-SN *g*-band, and ASAS-SN *V*-band, the few data points detected above the mean RMS and standard deviation of the controls and RMS of the data have large uncertainties and are thus consistent with the background. This suggests that there is no evidence of stochastic variability in these bands. Using the mean RMS of the ground-based optical light curves (see [Table A1](#)), we adopt an upper limit on the variability in these bands of $\lesssim 1.0 \times 10^5$ L$_\odot$.

To further investigate any long-term trends in the luminosity and search for coherent variability, we also fit the changes in the band luminosities with a simple linear function $L(t) = At + B$. For completeness, we also perform the same fit to the comparison sample. We find that slopes are both positive and negative across the bands and have a mean slope of $\sim 7.5 \times 10^2$ L$_\odot$ yr$^{-1}$. All bands have a slope that is consistent with zero and/or comparable to the mean slope of the control samples. This suggests that there is no evidence of a long-term variability during the final years before explosion down to a luminosity of $\lesssim 7.5 \times 10^2$ L$_\odot$ yr$^{-1}$ in these bands. This is consistent with that found by Neustadt et al. ([2023](#)), Hiramatsu et al. ([2023a](#)) and Dong et al. ([2023](#)).

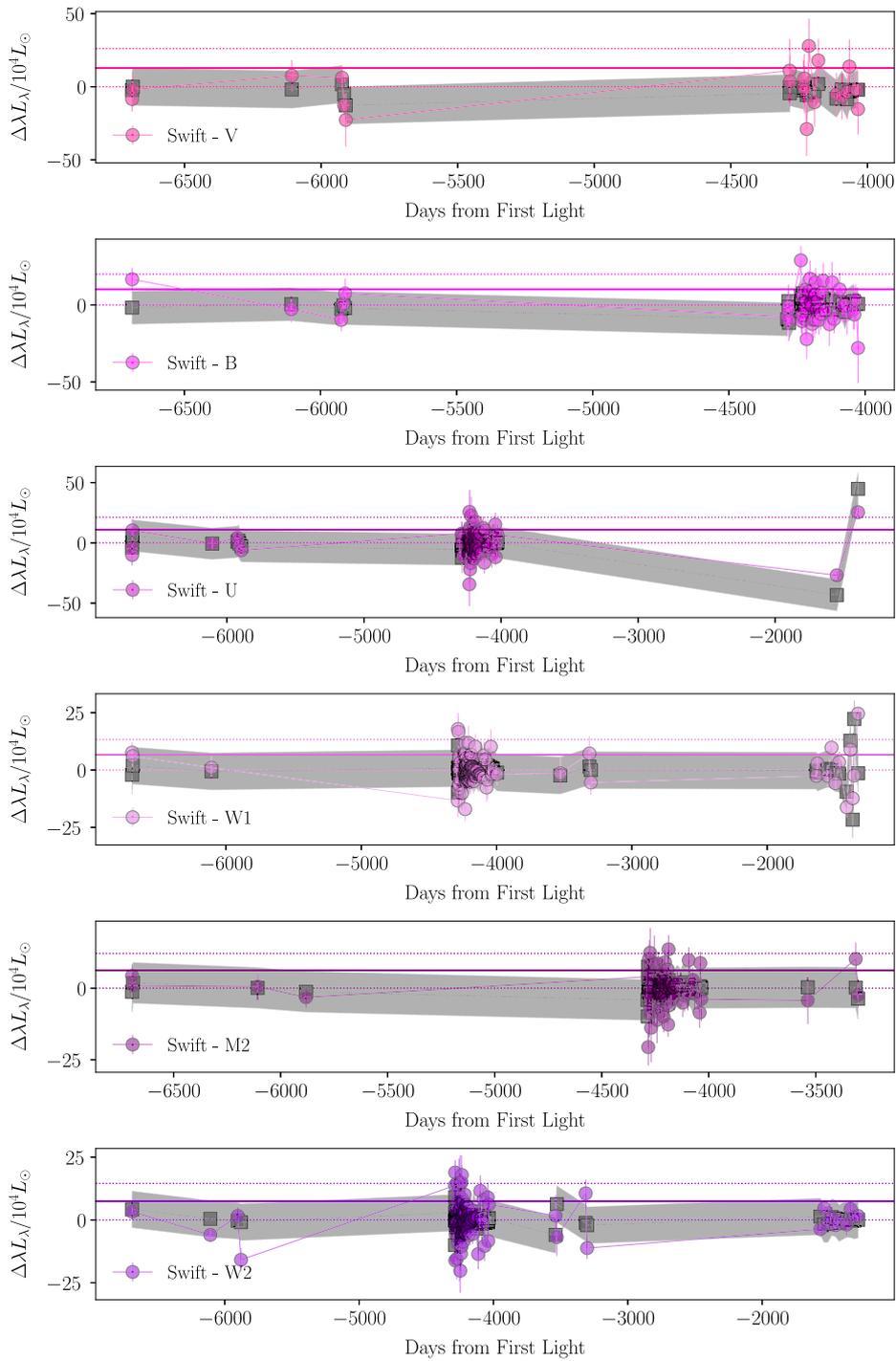

**Figure 7.** The peak-to-peak luminosity changes of the pre-SN differential luminosity ($\Delta \lambda L_\lambda$) of the SN2023ixf progenitor as observed in the *Swift* UVOT filters (solid coloured circles). The solid horizontal lines correspond to the root mean square of the peak-to-peak luminosity of our pre-explosion light curves, while the dotted lines correspond to the $1\sigma$ scatter. The grey squares correspond to the mean of the peak-to-peak luminosity changes of our comparison sample, while the shaded grey regions correspond to the standard deviation of this mean. The observed scatter in the luminosity of SN2023ixf's progenitor is consistent with the comparison sample, indicating no pre-SN variability of SN2023ixf at these wavelengths.

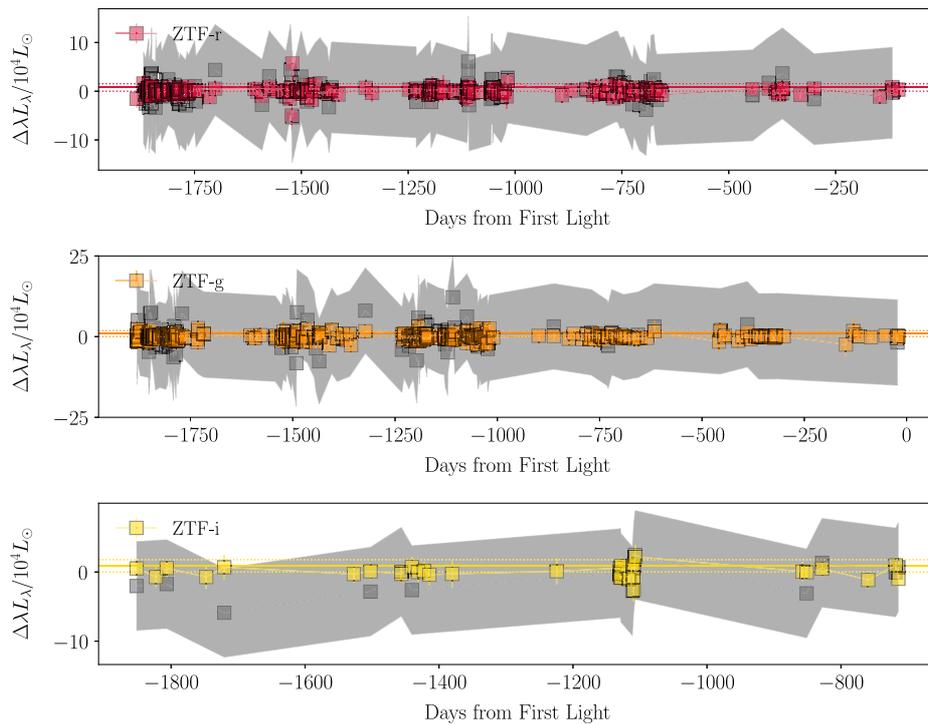

**Figure 8.** The peak-to-peak luminosity changes of the pre-SN differential luminosity ($\Delta\lambda L_\lambda$) of the SN2023ixf progenitor as observed in the ZTF filters. See Fig. 7 for more details.

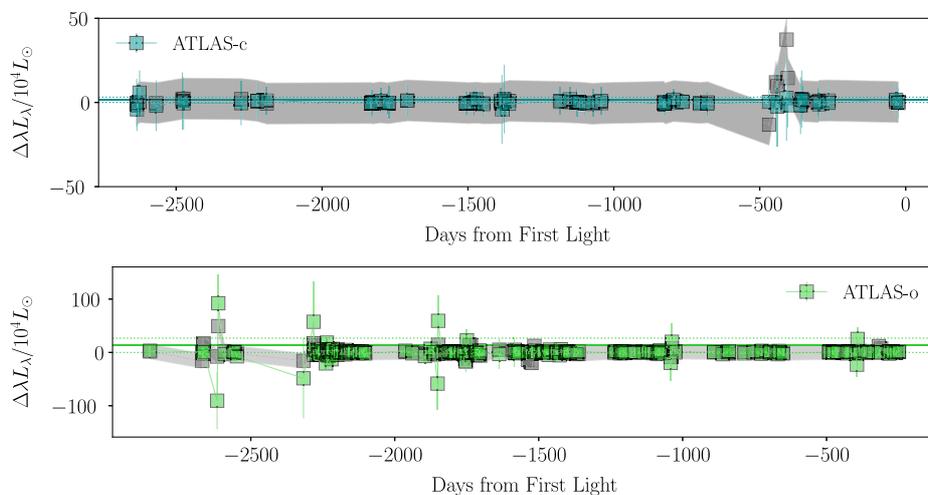

**Figure 9.** The peak-to-peak luminosity changes of the pre-SN differential luminosity ($\Delta\lambda L_\lambda$) of the SN2023ixf progenitor as observed in the ATLAS filters. See Fig. 7 for more details.

### 4.3 X-ray properties

Due to the proximity of M101, *Chandra*, *XMM-Newton*, and *Swift* have all observed SN2023ixf and the surrounding field in detail. As such, this provides us with the opportunity to place constraints not only on the X-ray emission associated with SN2023ixf as it evolves but also allows us to place constraints on the pre-explosion properties of the progenitor.

In Fig. 11, we show the X-ray light curve from the individual *Chandra* and *XMM-Newton* observations, as well as from the combined *Swift* observations. For the *Swift* observations, we also show hardness ratios for the emission detected after discovery.

In Fig. 12, we show our *Swift* X-ray spectrum. To estimate the X-ray luminosity, we convert our extracted count rate from both *Swift*, *Chandra*, and *XMM-Newton* into a flux using WebPIMMS[f] and we assume an absorbed bremsstrahlung model redshifted to M101 that has a temperature of $\Gamma = 35$ keV and a column density of $N_H \sim 3.9 \times 10^{22}$ cm$^{-2}$ (see Section 4.3.2 for more discussion about the spectral properties).

[f]https://heasarc.gsfc.nasa.gov/cgi-bin/Tools/w3pimms/w3pimms.pl.

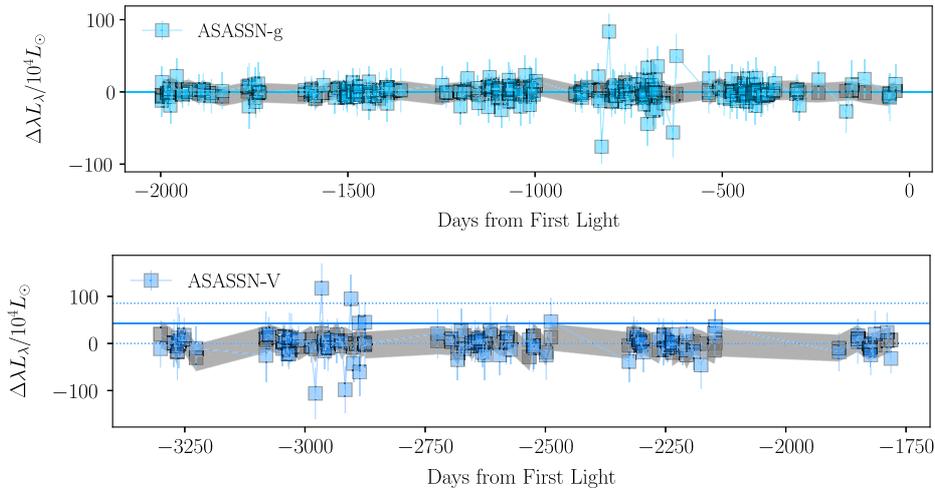

**Figure 10.** Differential luminosity of the SN2023ixf progenitor as observed in the ASAS-SN filters. See Fig. 7 for more details.

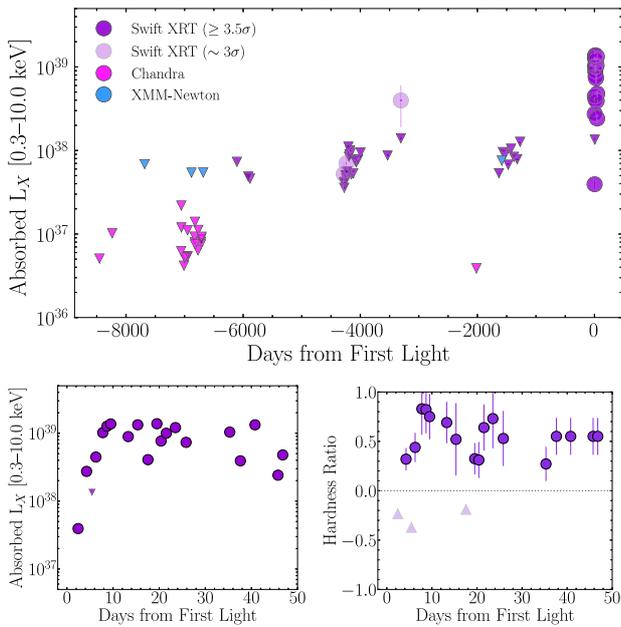

**Figure 11.** *Upper:* Broadband (0.3–10.0 keV) X-ray light curve of SN2023ixf prior to explosion as seen by *Chandra, XMM-Newton,* and *Swift.* The down arrows correspond to $3\sigma$ (or $3.5\sigma$) upper limits and the lightly shaded data points correspond to $3\sigma$ X-ray detections which we argue are most likely associated with a nearby HMXB. *Lower left:* The broadband X-ray light curve of SN2023ixf post-explosion emission as seen by *Swift.* Note that the error bars are smaller than the plotter markers. *Lower right:* The hardness ratios of the post-explosion X-ray emission, with up arrows indicating the $3\sigma$ lower limits. Here, the hardness ratio is derived using $\mathrm{HR} = (\textit{Soft Counts} - \textit{Hard Counts})/(\textit{Soft Counts} + \textit{Hard Counts})$, where Soft Counts corresponds to the count rate in the 0.3–2.0 keV energy band and Hard Counts corresponds to the count rate in the 2.0–10.0 keV energy band. The emission is relatively hard, consistent with a high-temperature thermal component.

### 4.3.1 Pre-explosion

We find no significant evidence of X-ray emission arising from the location of SN2023ixf up to 8 500 days prior to discovery (see Fig. 11 upper panel). The most constraining limit comes from the available *Chandra* X-ray observations, which are nearly an order of magnitude deeper than the *XMM-Newton* and *Swift* limits presented in Fig. 11. As such, to place the deepest constraint on the X-ray emission at the location of SN2023ixf, we merge all *Chandra* data together and determine a $3\sigma$ count rate upper limit of $8 \times 10^{-5}$ counts/sec in the 0.3–10.0 keV energy range. All upperlimits are derived following the methodology outlined in Section 2.2 of Auchettl et al. (2017). Assuming our best-fit, absorbed-bremsstrahlung model (see Section 4.3.2), this gives us an absorbed flux (0.3–10.0 keV) of $2.1 \times 10^{-15}$ erg cm$^{-2}$ s$^{-1}$, which corresponds to an absorbed (unabsorbed) X-ray luminosity (0.3–10.0 keV) of $3.3 \times 10^{36}$ erg s$^{-1}$ ($5.7 \times 10^{36}$ erg s$^{-1}$). If we use the more recent, but shallower *Swift* observations that cover up to 6 000 days prior to explosion, we obtain a $3\sigma$ count rate upper limit of $2 \times 10^{-4}$ counts/s in the 0.3–10.0 keV energy range, which corresponds to an unabsorbed X-ray luminosity (0.3–10.0 keV) of $5.8 \times 10^{37}$ erg s$^{-1}$ assuming our best-fit, absorbed-bremsstrahlung model.

In the left panel of Fig. 4, we show the pre-explosion image from our merged *Chandra* observations, while in the centre panel of Fig. 4, we have the pre-explosion image created by merging all available pre-explosion *Swift* data.

We do note that after we merged all *Swift* observations taken prior to discovery (Fig. 4 middle panel), our analysis suggests the presence of possible weak ($\sim 3\sigma$) X-ray emission coincident with the location of the SN. This is apparent for the observations found between 3 200 and 4 300 days prior to discovery (see Fig. 11 upper panel). Using the same model as above, this emission has an absorbed X-ray luminosity (0.3–10.0 keV) between $4.8 \times 10^{37}$ erg s$^{-1}$ and $3.7 \times 10^{38}$ erg s$^{-1}$. However, we note that close to the location of SN2023ixf, our merged *Chandra* observations (Fig. 4 left panel) show the HMXBs CXO J140341.1+541903, CXO J140336.1+541924, and [CHP2004] J140339.3+541827 (black, cyan, and magenta crosses in Fig. 4, Evans et al. 2010; Mineo, Gilfanov, & Sunyaev 2012). Unfortunately, due to the spatial resolution of the *Swift* XRT, it is difficult to disentangle the emission from SN2023ixf and that associated with these nearby binary systems (see Fig. 4 middle panel). As such, we believe it likely that this emission is associated with these sources rather than pre-explosion X-ray emission from the SN itself.

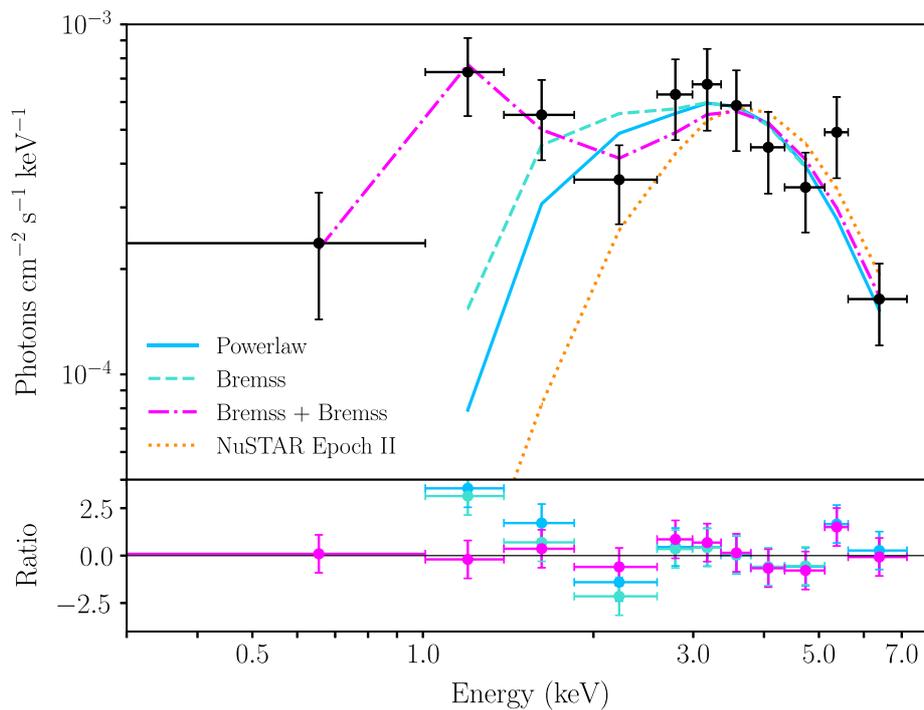

**Figure 12.** *Swift* XRT X-ray spectrum obtained by merging all observations taken within the first ∼46 days of first light (black data points). Shown as the magenta dot-dashed line is the best-fit, two-component, absorbed bremsstrahlung model. The teal dashed line is our best-fit, single-component, absorbed bremsstrahlung model, while the solid blue line is our best-fit, single-component, absorbed power law model. We have also plotted as an orange dotted line the best-fit model from Grefenstette et al. (2023) using their second epoch *NuSTAR* observations. Shown in the bottom panel are the residuals of our best-fit models.

### 4.3.2 Post-explosion

*Swift* began monitoring the evolution of SN2023ixf ∼0.67 days after discovery (MJD = 60083.73), or ∼1.57 days after first light (MJD = 60082.833). *Swift* did not detect soft X-ray (0.3–10.0 keV) emission until 4.25 days after first light (or 3.36 days after discovery). This is in contrast to the discovery of hard X-rays (> 3keV) at ∼3.9 days after first light using *NuSTAR* (Grefenstette et al. 2023). However, this is not surprising since the significantly enhanced column density measured by Grefenstette et al. (2023) at this time implies the soft X-rays would be absorbed. To place the strongest constraints on the emission during the first ∼4 days after first light, we merged the first 13 post-discovery *Swift* observations of SN2023ixf and from this calculated a 3σ upper limit to the count rate of $5 \times 10^{-4}$ counts/s in the 0.3–10.0 keV energy range.

At 4.25 days after first light, *Swift* detected X-ray emission with a significance of >4.5σ (see Fig. 4 lower left panel). This emission has continued to rise nearly an order of magnitude to its current peak count rate of $0.0094 \pm 0.002$ counts/s, corresponding to an absorbed X-ray (0.3–10 keV) luminosity of $(1.6 \pm 0.4) \times 10^{39}$ erg s$^{-1}$ assuming our best-fit bremsstrahlung model discussed below.

Unfortunately, due to the relatively short exposures of the *Swift* observations, we are unable to extract a spectrum from each individual observation. As such, we merged all available post-explosion *Swift* observations and extracted a spectrum to constrain the nature of this emission. This spectrum describes the soft X-ray emission arising from SN2023ixf in the first ∼46 days since first light and its best-fit spectral model discussed below is shown in Fig. 12.

To study the nature of this emission, we fit the 0.3–10.0 keV spectrum using an absorbed power law redshifted to the host galaxy (XSPEC model: TBABS*ZASHIFT*POWERLAW), similar to that done by Grefenstette et al. (2023). However, due to the emission seen at energies ≲2 keV, we find that our best-fit model with a column density of $N_H = (3.4^{+1.5}_{-1.1} \times 10^{22})$ cm$^{-2}$ and a photon index of $\Gamma = 1.6 \pm 0.5$ fails to fit the emission seen at these lower energies. However, if we instead fit the spectrum at energies ≳2 keV, our fit is consistent with that of Grefenstette (2023) using *NuSTAR* data.

Assuming this power law results from either X-ray synchrotron emission or inverse Compton (IC) scattering as is typical for X-ray emitting SNe (e.g. Chevalier & Fransson 2006), then the observed X-ray flux would be proportional to $\nu^{1-\Gamma}$, where $\nu$ is the frequency and $\Gamma$ is the photon index derived from our best-fit power law. As such, a value of $\Gamma = 1.6$ as derived above, would indicate a hard X-ray spectrum, which is consistent with that seen from other X-ray bright SNe (e.g. Chandra 2018).

However, for SNe that are interacting with a dense CSM, their fast shocks result in hard X-ray emission that is dominated by thermal bremsstrahlung emission with temperatures >1 keV (e.g. as seen in SN2014C, SN2010jl, and SN2006jd: Chandra et al. 2012, 2015; Katsuda et al. 2016; Margutti et al. 2017; Thomas et al. 2022; Brethauer et al. 2022). Since these high-temperature thermal models can mimic hard power laws, we also fit the spectrum using an absorbed, thermal-bremsstrahlung model (XSPEC model: TBABS*ZASHIFT*BREMSS). As with our best-fit power law model, our best-fit thermal model with a column density of $N_H = (2.4^{+0.7}_{-0.5} \times 10^{22})$ cm$^{-2}$ and a temperature of $kT = 33^{+50}_{-30}$ keV fails to fit the emission seen at lower energies (see Fig. 12). Nonetheless, the thermal model is a better fit than the single power law and is consistent with that found by Chandra et al. (2023b) using *Chandra* and by Grefenstette et al. (2023) using a second epoch of *NuSTAR* observations taken ∼10 days after first light.

In an attempt to explain this excess seen at lower energies, we tried adding either an absorbed bremsstrahlung model or an absorbed power law model to our best-fit bremsstrahlung model. This is similar to what was done for SN2010jl (Fransson et al. 2014; Chandra et al. 2015), whose *Chandra* and *Swift* observations showed the presence of an additional soft component ∼40–600 days after discovery, which the authors attributed to the cooling of the shock front. We find that an additional bremsstrahlung model significantly improves the fit. With this additional component, the spectrum is best fit (reduced $\chi^2 = 1.02$) with a column density of $N_{H,soft} = (0.9^{+0.5}_{-0.3} \times 10^{22})$ cm$^{-2}$ and a temperature of $kT_{soft} = 0.3 \pm 0.1$ keV for the low-energy (soft) component, and a column density of $N_{H,hard} = (3.9^{+1.2}_{-1.7} \times 10^{22})$ cm$^{-2}$ and a temperature of $kT_{hard} = 35^{+110}_{-27}$ keV for the high-energy (hard) component. This fit and its residuals are plotted in magenta in Fig. 12.

To verify whether this additional soft X-ray component can be attributed to SN2023ixf or instead arises from the nearby soft X-ray sources shown in Fig. 4, we extracted a merged *Swift* spectrum from the location of SN2023ixf prior to explosion. From this analysis, we find that the additional bremsstrahlung component required to accurately model the emission from SN2023ixf at energies less than 2 keV indeed arises from the supernova itself and does not arise from the complex X-ray environment in which SN2023ixf is situated.

To derive the fluxes and luminosities required for our analysis in Section 5, we use our best-fit column density ($N_{H,hard} = 3.9 \times 10^{22}$ cm$^{-2}$) and temperature ($kT_{hard} = 35$ keV) associated with the high-temperature component of our best-fit, two-component, bremsstrahlung model. This high temperature is thought to result from the expanding forward shock, meaning we can use this temperature to place constraints on the shock velocity and compare our results to those determined from optical spectra. To calculate the shock velocity, we use $v_s = [(16 k_B T_{sh})/(3 \mu m_H)]^{1/2}$, where $v_s$ is the shock velocity, $k_B T_{sh}$ is our shock temperature, $\mu = 0.604$ is the mean atomic weight, $k_B$ is Boltzmann's constant, and $m_H$ is the mass of hydrogen (e.g. Auchettl, Slane, & Castro 2014). Using this equation and our temperature from the high-energy component of our best-fit, two-component bremsstrahlung model, we find a shock velocity of $5\,440^{+5630}_{-2840}$ km s$^{-1}$ which is consistent with that found by Jacobson-Galan et al. (2023) using Hα absorption profiles and Grefenstette et al. (2023) using *NuSTAR* data.

## 5. Discussion

### 5.1 Lack of precursor variability from optical to X-rays

One way of constraining pre-explosion mass loss is to search for variable precursor emission from the progenitor. Evidence of end-of-life mass loss may also be seen in the form of precursor outbursts that are detected in the weeks to months before the main SN explosion (Ofek et al. 2014). These outbursts are thought to be associated with the different stages of nuclear burning (e.g. Fuller

2017; Shiode & Quataert 2014; Wu & Fuller 2021), instabilities in nuclear shell burning (e.g. Woosley et al. 2002), or interaction with a binary companion (e.g. Matsuoka & Sawada 2023).

Such an analysis has been performed for SN2023ixf in both the mid- and near-IR using *Spitzer* and ground-based observations, respectively (Kilpatrick et al. 2023; Jencson et al. 2023; Soraisam et al. 2023b), in some optical bands (Neustadt et al. 2023; Dong et al. 2023), and in the UV with GALEX (Flinner et al. 2023). Kilpatrick et al. (2023), Jencson et al. (2023), and Soraisam et al. (2023b) showed that the progenitor exhibited significant variability with a period of ∼1 000–1 200 days and an amplitude similar to the more luminous population of pulsating, dusty RSGs. However, Jencson et al. (2023) found no evidence for re-brightening due to eruptive, pre-SN outbursts predicted from early, post-SN observations (e.g. Jacobson-Galan et al. 2023), nor due to those expected from instabilities on the timescales of the final nuclear burning stages (e.g. Woosley et al. 2002). Similarly, Neustadt et al. (2023), Hiramatsu et al. (2023a) and Dong et al. (2023) found no evidence of variability in the optical, with Neustadt et al. (2023) deriving a limit of <100 L$_\odot$ yr$^{-1}$, suggesting that the progenitor did not undergo a luminous outburst within the 15 yr prior to discovery or it would have led to changes in the optical depth, making it detectable.

We search for evidence of possible pre-explosion variability/emission not only in the bands similar to that observed by Neustadt et al. (2023), Hiramatsu et al. (2023a) and Dong et al. (2023) but also using X-ray data from *Swift*, and UV/optical data from the *Swift* UVOT, as well as that observed using ATLAS, ZTF, and ASAS-SN. Here, we consider not only emission during the last year prior to discovery but also the possibility of pre-explosion variability spanning nearly 2 decades of observations.

Following the analysis presented in Johnson et al. (2017, 2018) and Neustadt et al. (2023) (see Section 4.2), we found no evidence of variability in optical to UV bands as detected by *Swift*, ASAS-SN, ZTF, and ATLAS in the ∼20 yr prior to explosion. This is consistent with that found by Neustadt et al. (2023), Hiramatsu et al. (2023a) and Dong et al. (2023). We derived a limit on the stochastic variability of $\lesssim 1 \times 10^5$ L$_\odot$ and on the long-term variability of $<7.5 \times 10^2$ L$_\odot$ yr$^{-1}$. These values are consistent with that obtained by Neustadt et al. (2023). In X-rays, we find no evidence of pre-explosion emission down to a limit of $5.8 \times 10^{37}$ erg s$^{-1}$ in the 16 yr (∼6 000 days) prior to explosion.

Neustadt et al. (2023) suggested that due to the heavy obscuration associated with the source, any short-lived outburst with a peak luminosity $\gtrsim 5 \times 10^6$ L$_\odot$ should lead to a detectable signature in the decade-long light curves, unless it lined up with a seasonal gap. Due to the extensive coverage by *Swift*, ASAS-SN, ATLAS, and ZTF prior to explosion, our results suggest that there was no bright, short-lived, pre-SN outburst within the ∼5 yr prior to explosion, similar to that found by Dong et al. (2023) and Hiramatsu et al. (2023a). In addition, our results for *Swift*, ASAS-SN, ATLAS, and ZTF in Table A1 require that any outburst had a luminosity $\lesssim 1 \times 10^5$ L$_\odot$. This is consistent with the suggestion by Neustadt et al. (2023), Dong et al. (2023), and Jencson et al. (2023) that there was no luminous, short-lived outburst within the 15 yr prior to explosion. In addition, the lack of variability detected by Hiramatsu et al. (2023a) is consistent with their dense, confined CSM models that take into account pre-existing dust surrounding the progenitor. However, these findings are in contrast to the prediction by Jacobson-Galan et al. (2023) that the dense CSM inferred from early spectra could have arisen if the progenitor experienced periods of enhanced mass loss during the final 3–6 yr before core collapse.

The lack of variability seen in these wavelengths may not be so surprising. Currently, there have been nearly 30 SNe that have exhibited pre-SN activity (e.g. Jacobson-Galán et al. 2022a; Strotjohann et al. 2021), but current surveys searching for precursors and SN progenitors (e.g. Johnson et al. 2017, 2018) suggest that not all SNe exhibit luminous precursors and at most, ∼2.5% of all core-collapse SNe or ∼25% of Type IIn SNe produce luminous, eruptive outbursts prior to explosion (Strotjohann et al. 2021). As such, it is possible that the formation of SN2023ixf's CSM results not from episodic mass loss, but rather from binary interactions (Matsuoka & Sawada 2023) that gave rise to its observed asymmetry (Smith et al. 2023; Vasylyev et al. 2023). Here, the collision of the winds from the two stars in the binary system would produce a high-density (and possibly asymmetric) CSM around the progenitor (e.g. Kochanek 2019).

## 5.2 Pre-explosion mass loss

Using pre- and post-explosion X-ray observations, it is possible to place constraints on the density of the CSM. This is because the mass that is ejected from the progenitor star during the pre-SN phase becomes the CSM with which the SN ejecta interacts following the explosion.

### 5.2.1 Pre-explosion X-ray constraints on the mass loss

The interaction of the shock with the CSM can power the resulting SN and produce both non-thermal X-rays and radio emission (e.g. Margutti et al. 2014, 2017). However, evidence of mass loss may also be seen as precursor outbursts in X-rays and at other wavelengths. Due to its proximity, the location of SN2023ixf has been extensively observed using *Chandra*, *Swift*, and *XMM-Newton* (see Fig. 4). Similar to our analysis of the pre-explosion optical and UV light curves (see Section 4.2), our pre-explosion X-ray analysis also finds no evidence of luminous precursor emission at higher energies.

Using our deep pre-explosion *Chandra* observation, we can place constraints on the pre-explosion mass loss. Since we find no evidence of outbursts that would be consistent with an eruption scenario, we assume that any precursor emission would have been generated by a wind with a mass-loss rate $\dot{M}$ and wind velocity $v_w$. Equation 30 from Matsumoto & Metzger (2022) allows us to estimate the mass-loss rate using the progenitor mass ($M_\star$), wind velocity ($v_w$), and precursor luminosity ($L_{pre}$) according to:

$$\dot{M} \simeq 1.5 \left( \frac{M_\star}{10 \ M_\odot} \right)^{-2} \left( \frac{v_w}{10^3 \ \mathrm{km \ s^{-1}}} \right)^{-2} \left( \frac{L_{pre}}{10^{40} \ \mathrm{erg \ s^{-1}}} \right)^3 M_\odot \ \mathrm{yr^{-1}}.$$

Adopting a progenitor mass between 11 and 20 $M_\odot$ (Kilpatrick et al. 2023; Jencson et al. 2023; Soraisam et al. 2023b), a wind velocity of 50 km s$^{-1}$, and using for $L_{pre}$ our unabsorbed 3σ upper limit to the X-ray emission from *Chandra* observations that span >6 000 days (>18 yr) prior to explosion ($L_X < 5.7 \times 10^{36}$ erg s$^{-1}$), we get $\dot{M} = (3-9) \times 10^{-8} M_\odot$ yr$^{-1}$. If we use the more recent *Swift* upper limit derived from merging all available *Swift* pre-explosion observations that cover <6 000 days (<16 yr) prior to explosion ($L_X < 5.8 \times 10^{37}$ erg s$^{-1}$), we get $\dot{M} = (3-10) \times 10^{-5} M_\odot$ yr$^{-1}$.

The mass-loss rate derived using the more recent *Swift* constraint (covering < 16 yr prior to explosion) is consistent with the mass loss expected for normal RSGs (Beasor et al. 2020) and that derived by Neustadt et al. (2023) and Jencson et al. (2023), who modelled the IR SED of the progenitor assuming it is surrounded by a dusty CSM. However, it is lower than that derived by Jacobson-Galan et al. (2023) from early spectra and Soraisam et al. (2023c) who used the mass-loss prescription of Goldman et al. (2017) and a SED-independent luminosity for the progenitor. As our mass-loss rate from deep *Chandra* observations spanning >18 yr prior to explosion is even more constraining, this could suggest that the mass loss may have increased over the two decades prior to explosion. This is consistent with the results of Jencson et al. (2023), whose findings suggest that the progenitor had a steady but enhanced wind that developed over the final decade or more prior to explosion.

### 5.2.2 Post-explosion X-ray constraints on the mass loss

Post-explosion X-ray observations can also be used to place deep constraints on the properties of the CSM for Type Ia (e.g. Margutti et al. 2012), Type Ibc (e.g. Drout et al. 2016), and Type Ib/IIn SNe such as SN2014c (e.g. Margutti et al. 2014). In these papers, they follow the formalism that within the first ∼ month of evolution, the X-ray emission is dominated by IC scattering that results in photons from the photosphere being upscattered to X-ray energies by relativistic electrons from the expanding SN shock front (Björnsson & Fransson 2004; Chevalier & Fransson 2006; Margutti et al. 2012, 2017).

If the progenitor lost material at a constant rate $\dot{M}$, which seems to be consistent with the results presented in Section 5.2 and those derived by Jencson et al. (2023), we can use the formalism presented in Equations A7 and A8 of Margutti et al. (2012) and Margutti et al. (2018b) to calculate the wind density ($\dot{M}/v_w$). We can also use the fact that *Swift* observations of SN2023ixf taken within the first ∼3.3 days after first light showed no soft X-rays. Merging these observations, we derive a 3σ upper limit to the unabsorbed luminosity in the 0.3–10.0 keV energy band of $7.2 \times 10^{38}$ erg s$^{-1}$.

To calculate the corresponding IC emission, we assume the following: that the ejecta has a density profile that follows $\rho \propto R^{-n}$ with $n \sim 9$ as appropriate for compact progenitors (Matzner & McKee 1999), that the shock-accelerated electrons are best described by a power-law distribution with an index of 3, that 1% of the post-shock energy density goes in relativistic electrons (Reynolds et al. 2021), that the shock velocity depends on the wind velocity, explosion energy, ejecta mass, and CSM density which is described by $\rho_{CSM} = \dot{M}/4\pi v_w R^2$, and that $L_{X-ray} \propto L_{bol}$, where $L_{bol}$ is the bolometric luminosity of the source.

Using an explosion energy of $10^{51}$ erg and an ejecta mass of $M_{ej} = 1.8 \ M_\odot$ as estimated in Section 4.1, we find that the lack of soft X-rays during the first ∼3.3 days after first light implies a mass-loss rate of $\dot{M} \lesssim 5 \times 10^{-4} \ M_\odot$ yr$^{-1}$, assuming a wind velocity of 50 km s$^{-1}$. Using our derived mass-loss rate, and Equations A3 and A7 from Margutti et al. (2018b), we estimate that this low-density environment occurs out to a distance of $R < 3.7 \times 10^{15}$ cm.

As *NuSTAR* observations showed that the source was X-ray bright in the hard X-ray band (Grefenstette et al. 2023), it is possible that SN2023ixf was emitting X-rays much earlier than the first *Swift* detection at 4.26 days after first light, but that the exposure time and sensitivity of the *Swift* observations were not sufficient to detect the highly-absorbed, high-temperature emission from the source. As such, if we perform the calculation above but instead use the time of the first *Swift* observation (1.57 days after first light) and its corresponding bolometric luminosity, we get $\dot{M} < 3.1 \times 10^{-4} \, M_\odot \, \mathrm{yr}^{-1}$ and $R < 1.8 \times 10^{15}$ cm.

These results and those derived above are consistent with that calculated using our pre-explosion constraints and those published in the literature, including the mass-loss rates from Grefenstette et al. (2023) using *NuSTAR* data and from Soraisam et al. (2023c) using the empirical period-luminosity-based mass-loss prescription from Goldman et al. (2017). In addition, our shock radius is consistent with those derived by Grefenstette et al. (2023), Smith et al. (2023), and Jacobson-Galan et al. (2023). However, our calculated mass-loss rate differs from other estimates in the literature, including that of Neustadt et al. (2023, $\sim 10^{-5} \, M_\odot \, \mathrm{yr}^{-1}$), Niu et al. (2023, $1 \times 10^{-5} \, M_\odot \, \mathrm{yr}^{-1}$), Bostroem et al. (2023, $10^{-3} - 10^{-2} \, M_\odot \, \mathrm{yr}^{-1}$), Jacobson-Galan et al. (2023, $10^{-2} \, M_\odot \, \mathrm{yr}^{-1}$), and Hiramatsu et al. (2023a, $0.1 - 1 \, M_\odot \, \mathrm{yr}^{-1}$ (1–2 yr before explosion) and $0.1 - 1 \, M_\odot \, \mathrm{yr}^{-1} \sim$ (0.7-0.4 yr before explosion)). These discrepancies may be attributed to a number of factors including the chosen method or adopted parameters. For example, Niu et al. (2023) used the mass-loss rate prescription from Beasor & Davies (2016) which relies on the progenitor mass, while Bostroem et al. (2023) and Jacobson-Galan et al. (2023) compared the spectroscopic evolution of SN2023ixf to a grid CMFGEN models that were sensitive to the adopted wind velocity, $v_w$.

We can also use our best-fit hydrogen column density ($N_{H,hard} = (3.9^{+1.2}_{-1.7} \times 10^{22})$ cm$^{-2}$) to place a constraint on the mass loss, which yields a value consistent with Grefenstette et al. (2023). Equation 4.1 from Fransson, Lundqvist, & Chevalier (1996) gives

$$N_H = (2.1 \times 10^{22}) \left( \frac{1}{\mathrm{s}^{-1}} \right) \left( \frac{v_w}{10 \, \mathrm{km \, s^{-1}}} \right)^{-1} \left( \frac{\dot{M}}{10^{-5} \, M_\odot} \right) \times$$
$$\left( \frac{v_s}{10^4 \, \mathrm{km \, s^{-1}}} \right)^{1-s} \left( \frac{t}{8.90 \, \mathrm{days}} \right)^{1-s} \mathrm{cm}^{-2},$$

where $N_H$ is the column density derived in Section 4.3, t is the time at which this column density was measured, $\dot{M}$ is the mass-loss rate, $v_s$ is the shock velocity as derived in Section 4.3, $v_w = 50$ km s$^{-1}$ is the wind velocity, and $s = 2$ is the index of the density profile. We find that the mass-loss rate of the progenitor was $(1.6^{+0.9}_{-1.0} \times 10^{-4}) \, M_\odot \, \mathrm{yr}^{-1}$, which is similar to both Grefenstette et al. (2023) and what we derived above.

Our derived shock radius suggests that the CSM of SN2023ixf is relatively compact. This is consistent with Grefenstette et al. (2023), who measured a rapidly decreasing column density between their observations but also with the disappearance of CSM interaction signatures within 8 days of discovery (e.g. Jacobson-Galan et al. 2023; Smith et al. 2023) and the non-detection at millimetre wavelengths (Berger et al. 2023a).

Assuming a constant shock velocity, we can use the ratio of the shock velocity ($v_{shock}$) to the CSM velocity (or the wind velocity, $v_{wind}$) and the time of observation ($t_{obs}$) to place constraints on the pre-SN ejection time of the CSM ($t_{ejection}$) according to $t_{ejection} > t_{obs}(v_{shock}/v_{wind})$ from Dickinson et al. (2023). Using our derived mass-loss rate, the shock velocity derived in Section 4.3 and a wind velocity of 50 kms$^{-1}$, we find that SN2023ixf exhibited a mass-loss episode at least $> 0.5 - 1.5$ ($v_w/50$ km s$^{-1}$) yr prior to explosion. This is consistent with Jacobson-Galan et al. (2023), Kilpatrick et al. (2023), Jencson et al. (2022), and Smith et al. (2023), who suggested that the progenitor experienced enhanced mass loss >1 yr prior to explosion.

### 5.3 X-ray constraints on the CSM

In Fig. 12, we present our merged *Swift* X-ray spectrum with the best-fit models discussed in Section 4.3 and a model using the best-fit parameters ($N_H = 5.6 \times 10^{22}$ cm$^{-2}$ and $kT = 34$ keV) from epoch two ($\sim 11$ days after first light) associated with the early *NuSTAR* data from Grefenstette et al. (2023). While both our single-component models and the *NuSTAR* model describe the data equally well for energies > 3 keV, these models underestimate the emission at energies < 2 keV. We find that an additional low-temperature, absorbed bremsstrahlung component is able to reproduce the emission seen at energies < 2 keV.

Similar to the well-studied Type IIn SN2010jl (Fransson et al. 2014; Chandra et al. 2015), we attribute this soft component to the cooling of the forward shock. Using the mass-loss constraint derived in Section 5.2.2 and the shock velocity derived from our high-temperature, bremsstrahlung component (see Section 4.3.2), we can calculate the cooling time. To do this we use Equation 6 from Fransson et al. (2014):

$$t_{cool} = 26.6 \left( \frac{\dot{M}}{0.1 \, M_\odot \mathrm{yr}^{-1}} \right)^{-1} \left( \frac{v_w}{100 \, \mathrm{km \, s^{-1}}} \right)$$
$$\times \left( \frac{V_s}{3\,000 \, \mathrm{km \, s^{-1}}} \right)^3 \left( \frac{t}{\mathrm{years}} \right)^{1.46} \mathrm{days},$$

where $\dot{M}$ is the mass-loss rate, $v_w$ is the wind velocity, $V_s$ is the shock velocity, and $t$ is the time at which the mass-loss rate was measured. Assuming $v_w = 50$ km s$^{-1}$, $\dot{M} = 5 \times 10^{-4} \, M_\odot \, \mathrm{yr}^{-1}$, $V_s = 5441$ km s$^{-1}$, and $t \sim 3.3$ days, we obtained a cooling time of $t_{cool} \sim 17$ days. Thus, it is possible that the forward shock is indeed cooling, giving rise to this additional component.

If we assume that the shock is propagating through a wind characterised by $\rho \propto R^{-2}$, we would expect to see a decrease in the measured column density as the shock evolves, provided that the vast majority of the measured column density arises from CSM. As the Galactic absorption is much less than the measured column density derived in our analysis ($N_{H,gal} = 7.9 \times 10^{20}$ cm$^{-2}$, HI4PI Collaboration et al. 2016), this is not an unreasonable assumption. In fact, the decrease in column density has been seen by both Grefenstette et al. (2023) and Chandra et al. (2023b), who found a column density of $N_H = 26 \times 10^{22}$ cm$^{-2}$ ~4 days after first light, which decreased to $N_H = 5.6 \times 10^{22}$ cm$^{-2}$ on day ~11 and then $N_H \sim 3.2 \times 10^{22}$ cm$^{-2}$ on day ~14. Combined with our constraint of $3.9 \times 10^{22}$ cm$^{-2}$ by ~46 days, we find that the column density is decreasing following a simple power law with an index of ~1.5.

This gradient is much steeper than expected assuming a steady, spherically symmetric wind (i.e. an index of 2). This could suggest that the source underwent either variable mass loss from the progenitor before explosion or the progenitor is surrounded by an intrinsically asymmetric CSM geometry (Fransson et al. 1996) such as that seen in other interacting SN e.g. SN2006jd, SN2010jl, and SN1998S; Chandra et al. 2012; Fransson et al. 2014; Leonard et al. 2000). The lack of pre-explosion outbursts (see Section 4.2, Neustadt et al. 2023, Jencson et al. 2023, Dong et al. 2023), and the evolution of the spectral features found in the high resolution spectra of SN2023ixf during its first week suggest that the CSM is likely asymmetric (e.g. Smith et al. 2023), supporting the case that the observed decay in column density is largely due to the geometry of the CSM.

We can also derive the mass of the CSM swept up by the forward shock using the following (Chandra 2018; Margalit,

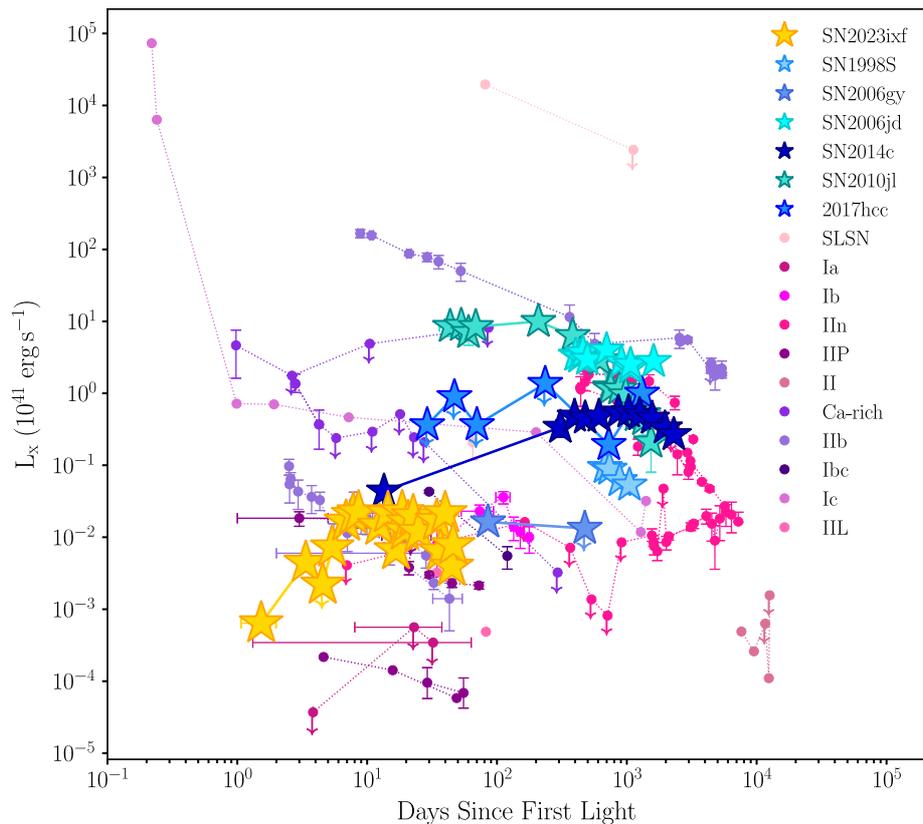

**Figure 13.** Unabsorbed X-ray luminosities (in units of $10^{41}$ erg s$^{-1}$) for SN2023ixf assuming a thermal bremsstrahlung model (0.3–10 keV, yellow stars) and other SNe detected in X-rays as a function of the time since first light (MJD=60082.83). Stars represent Type IIn SNe, while arrows indicate the measurement is an upper limit. For those SNe that are not Type IIn, the marker colour gives the SN type. *Data Source and energy band:* SN1970G (Immler & Kuntz 2005, (0.3–2 keV)), SN1978K (Chandra 2018, (0.3–8 keV)), SN1980K (Schlegel 2001, (2 keV)), SN1986J (Houck et al. 1998, (0.5–2.2 keV)), SN1988Z (Schlegel & Petre 2006, (0.2–2 keV)), SN1993J (Chandra et al. 2009, (0.3–8 keV)), SN1994W (Chandra 2018, (0.3–8 keV)), SN1995N (Chandra et al. 2005, (0.1–10 keV)), SN1996cr (Chandra 2018, (0.5 - 2 keV)), SN1998bw (Kouveliotou et al. 2004, (0.3–10 keV)), SN1998S (Pooley et al. 2002, (2–10 keV)), SN1999em (Schlegel 2001, (2 keV)), SN1999gi (Schlegel 2001, (2 keV)), SN2003bg (Soderberg et al. 2006, (0.5–8 keV)), SN2004et (Misra et al. 2007, (0.5–8 keV)), SN2005kd (Chandra 2018, (0.3–8 keV)), SN2005ip (Katsuda et al. 2014, (0.2–10 keV)), SN2006bp (Immler et al. 2007, (0.2–10 keV)), SN2006gy (Chandra 2018, (0.5–2 keV)), SN2006jc (Immler et al. 2008, (0.2–10 keV)), SN2006jd (Chandra 2018, (0.3–8 keV)), SCP06F6 (Levan et al. 2013, (0.2–10 keV)), SN2008ax (Roming et al. 2009, (0.2–10 keV)), SN2010jl (Chandra et al. 2015, (0.2–10 keV)), SN2011dh (Soderberg et al. 2012, (0.3–8 keV)), SN2011fe (Margutti et al. 2012, (0.5–8 keV)), SN2014c (Brethauer et al. 2022, (0.3–100 keV)), PTF12dam (Margutti et al. 2018a, (0.3–10 keV)), SN2017hcc (Chandra et al. 2022, (0.3–10 keV)), SN2018gk (Bose et al. 2021, (0.3–10 keV)), SN2019ehk (Jacobson-Galán et al. 2020, (0.3–10 keV)), SN2021gno (Jacobson-Galán et al. 2022b, (0.3–10 keV)).

Quataert, & Ho 2022): $M_{CSM} = \dot{M} R_s (v_w / 10 \text{ km s}^{-1})^{-1}$. During the first $\sim 46$ days, the shock had swept up $\sim 0.04$–0.07 $M_\odot$ of CSM. This is consistent with that estimated using optical spectra (Jacobson-Galan et al. 2023; Bostroem et al. 2023).

### 5.4 Comparison with other X-ray bright SNe

SNe that explode in dense CSM are expected to be dominated by thermal X-ray emission from the forward shock with temperatures on the order of 10 to a few 10s of keV (e.g. see review by Chandra 2018). The increase in X-ray emission seen from SN2023ixf indicates that its shock is interacting with a dense, hydrogen-rich CSM that was recently ejected by the progenitor prior to its death. This is consistent with our results presented in Section 4.3.2 and the growing body of evidence that pre-explosion mass loss occurs in the progenitors of Type IIn-like SNe through either binary interaction or episodic mass loss (e.g. see review by Smith 2014).

In Fig. 13, we plot our X-ray luminosity as a function of time for SN2023ixf assuming a thermal bremsstrahlung model and compare it to a sample of other X-ray bright SNe. The rise in the X-ray emission of SN2023ixf (Fig. 11, lower left panel) has been seen in a number of interacting SNe including Type IIn SN2010jl (Chandra et al. 2015) and SN2006jd (Chandra 2018), and the Type Ib SN2014c that transitioned into a strongly interacting Type IIn (Margutti et al. 2017; Thomas et al. 2022). However, the timescales to peak X-ray brightness for these other events are much longer ($\sim$100 days) compared to the $\sim$10 days it took for the soft X-rays from SN2023ixf to attain their current peak. This difference in rise time is likely due to the location of the CSM, with the CSM surrounding SN2023ixf likely more compact compared to the other SNe that show interaction or which emit X-rays (Fig. 13). This rise in X-ray brightness is also likely a result of the decrease in shock temperature as it sweeps and interacts with more material over time. This causes the X-ray emission to shift from higher to lower X-rays energies. Such behaviour is consistent with the decrease in shock temperature seen by NuSTAR (Grefenstette et al. 2023). The flattening in the X-ray light curve seen after $\sim$10 days (See Fig. 11 lower left panel) suggests that the X-ray emission is currently powered by CSM interaction. This behaviour has been seen in both SN2014c and other Type IIn SNe with well-sampled X-ray light curves (e.g. SN2010jl Chandra et al. 2015, SN2006jd Chandra et al. 2012 and see Fig. 13).

Compared to other Type IIn SNe, both the bolometric luminosity (see Fig. 6) and the X-ray luminosity of SN2023ixf (see Fig. 13) are currently a few orders of magnitude less than that seen for strongly interacting SNe such as SN2010jl (Chandra et al. 2015), SN2006jd (Chandra et al. 2012), and SN2005kd (Dwarkadas et al. 2016), and is more consistent in luminosity with that seen from Type IIn SN2006gy (Smith et al. 2007), SN1998S (Pooley et al. 2002), SN2017hcc (Chandra et al. 2022), or some Type IIP, Type IIb, or Type Ib SNe. The low X-ray luminosity of SN2023ixf is rather peculiar given: that Type IIn SNe tend to be much brighter in X-rays than other core-collapse SNe (e.g. Chandra 2018), that the shock appears to be interacting with dense material based on flash ionisation features (e.g. Jacobson-Galan et al. 2023; Smith et al. 2023; Jencson et al. 2023), the constraints on the CSM both in this work and others in the literature, the enhanced mass-loss rates derived in this and in other works, and the large column densities relative to the Galactic column density measured here and by NuSTAR (Grefenstette et al. 2023) and Chandra (Chandra et al. 2023b).

Similar to the Type IIn SN SN2017hcc, which exhibited weak X-ray and radio emission but bright IR emission, it is possible that SN2023ixf's low-luminosity X-ray emission arises from an asymmetric CSM. In this case, the conversion of kinetic energy is inefficient in some directions, causing a lower X-ray flux. Another possibility is that the X-ray emission is suppressed due to instabilities at the shock front or due to the hot post-shock gas driving weak shocks into the colder surrounding material, transferring energy before it can be radiated (e.g. Steinberg & Metzger 2018). As a number of studies in the literature suggest that the CSM is asymmetric (e.g. Smith et al. 2023; Berger et al. 2023b; Vasylyev et al. 2023), it is likely the lower X-ray luminosity is a signature of the CSM distribution.

We can expect that once the X-ray luminosity begins to fade, it will decay following $t^{-1}$ if the CSM was formed via a steady wind whose density decreases as $r^{-2}$ (e.g. Chevalier 1998; Dwarkadas & Gruszko 2012). However, it has been shown that a number of SNe that are known to interact with a dense CSM deviate from this behavior (see e.g. Dwarkadas & Gruszko 2012; Ross & Dwarkadas 2017; Chandra 2018). This could result from either a CSM that does not follow $r^{-2}$ (Salamanca 2003; Dwarkadas 2011; Dwarkadas & Gruszko 2012), or from the fact that our X-ray instruments observe in a narrow X-ray band (Dwarkadas & Gruszko 2012) and thus our bolometric corrections are not capturing the full emission. As our analysis suggests that the CSM was formed from low-luminosity, episodic mass loss or from binary interaction, we expect that the emission will decay following a power law that deviates from this behaviour. As such, it will be critical to continue monitoring the X-ray evolution of SN2023ixf as it evolves to further probe the mass-loss history of its stellar progenitor.

## 6. Summary and conclusions

We have presented a comprehensive study of the pre-explosion UV, optical, and X-ray properties, and post-explosion X-ray emission of the nearby Type IIn SN SN2023ixf as detected by Swift, ASAS-SN, ATLAS, ZTF, Chandra, and XMM-Newton. Here, we analysed data from nearly two decades prior to explosion and

∼50 days post explosion, focusing on the pre-explosion variability, presence of X-ray signatures prior to explosion and the evolution of the luminous soft X-ray emission. In summary:

1. Using nearly two decades of data, we showed that the progenitor of SN2023ixf exhibited no significant evidence of pre-explosive/precursor activity in the optical, UV, and X-rays as detected by Swift, ASAS-SN, ZTF, ATLAS, Chandra, and XMM-Newton. This is consistent with that found by Neustadt et al. (2023)), Hiramatsu et al. (2023a) and Dong et al. (2023). Our analysis suggests that any evidence of precursor activity in optical to UV was characterised by a luminosity ≲$1 \times 10^5 L_\odot$, consistent with Neustadt et al. (2023), Jencson et al. (2023), and Dong et al. (2023) who suggested that there was no bright, short-lived outburst within the last ∼ decades prior to explosion.

2. Extensive, serendipitous monitoring of the location of SN2023ixf by Chandra, Swift, and XMM-Newton prior to explosion shows no evidence of luminous precursor emission down to $5.7 \times 10^{36}$ erg s$^{-1}$ based on Chandra observations spanning more than 18 yr prior to the explosion. Using Swift observations from the progenitor's final 16 yr, we instead obtain a value of less than $5.8 \times 10^{37}$ erg s$^{-1}$ and derive a mass-loss rate of $(3-10) \times 10^{-5} M_\odot$ yr$^{-1}$, which is consistent with those rates derived by Neustadt et al. (2023) and Jencson et al. (2023).

3. Using the extensive set of Swift XRT observations of SN2023ixf that have been taken since its discovery, we find that Swift did not detect any soft X-ray emission down to an unabsorbed luminosity of $7.2 \times 10^{38}$ erg s$^{-1}$ within the first ∼3.3 days after first light. Assuming that the emission is dominated by IC scattering, similar to that done for other SNe (e.g. Margutti et al. 2012, 2014), we derive a mass-loss rate of ≲$5 \times 10^{-4} M_\odot$ yr$^{-1}$ and a radius of $R < 3.7 \times 10^{15}$ cm for the CSM. Our analysis suggests that the progenitor underwent a mass-loss episode at least $> 0.5 - 1.5(v_w/50$ km s$^{-1})$ yr prior to explosion.

4. By merging the available Swift XRT observations, we find that the emission over the first ∼50 days is best described by an absorbed, two-temperature component bremsstrahlung model. Here, the hard component has a temperature of ∼35 keV and a column density of $3.9 \times 10^{22}$ cm$^{-2}$, consistent with that found using NuSTAR (Grefenstette et al. 2023). The soft component has a temperature of ∼0.3 keV and a column density of $0.9 \times 10^{22}$ cm$^{-2}$, which we suggest results from the forward shock cooling, similar to that seen in SN2010jl. We also derive a swept up mass of 0.04–0.07 $M_\odot$ for the CSM.

5. Similar to other interacting SNe, we find that the X-ray emission of SN2023ixf has risen to peak brightness and is now plateauing. However, the rise to peak was much faster than other interacting SNe, while the peak luminosity is nearly a few orders of magnitude less than that seen for strongly interacting SNe such as SN2010jl or SN2006jd. This is peculiar considering the fact that SN2023ixf showed evidence of flash ionisation features and enhanced mass-loss rates. We suggest that the low X-ray luminosity may be a natural consequence of an asymmetric CSM, similar to that of the Type IIn SN 2017hcc and consistent with the findings of Smith et al. (2023), Berger et al. (2023b), and Vasylyev et al. (2023).

Due to its proximity and extensive pre- and post-explosion multiwavelength coverage, SN2023ixf provides a unique opportunity to not only understand the time-dependent mass loss, variability and CSM formation associated with the final stages of red supergiant evolution but also the possible effects of binarity on these systems. The continued monitoring of the electromagnetic emission (in addition to neutrinos and gravitational waves) from SN2023ixf will provide some of the strongest constraints on the mechanism associated with stellar collapse of a massive progenitor seen in the last few decades.

**Acknowledgement.** We thank C. Kochanek for his insight and advice.

**Funding statement.** Parts of this research were conducted by the Australian Research Council Centre of Excellence for Gravitational Wave Discovery (OzGrav), through project number CE170100004.

LAL acknowledges support by the Simons Foundation and the Heising-Simons Foundation. This work was performed in part at the Simons Foundation Flatiron Institute's Center for Computational Astrophysics during LAL's tenure as an IDEA Scholar.

JFB was supported by National Science Foundation grant No. PHY-2310018.

**Data availability.** Not applicable.

## Appendix

In Table A1, we present the results of our variability analysis as outlined in Section 4.2.

**Table A1.** Variability Limits for SN2023ixf in various bands.

| | | Variability ($10^4$ L$_\odot$) | | |
|---|---|---|---|---|
| | | Progenitor | Control sample | Control sample |
| Instrument | Band | RMS | Mean RMS | $\sigma$ |
| *Swift* | V | 26.23 | 4.67 | 12.72 |
| *Swift* | B | 17.76 | 3.51 | 10.68 |
| *Swift* | U | 10.93 | 8.03 | 12.92 |
| *Swift* | W1 | 8.45 | 4.35 | 8.00 |
| *Swift* | M2 | 8.44 | 2.11 | 7.06 |
| *Swift* | W2 | 10.72 | 2.31 | 7.24 |
| ZTF | r | 0.75 | 2.29 | 9.33 |
| ZTF | g | 1.02 | 3.39 | 13.26 |
| ZTF | i | 0.92 | 2.28 | 6.42 |
| ASASSN | g | 23.4 | 4.42 | 13.25 |
| ASASSN | V | 27.7 | 11.1 | 28.2 |
| ATLAS | c | 2.43 | 6.28 | 12.2 |
| ATLAS | o | 13.4 | 6.05 | 14.0 |